\documentclass[amsmath,amssymb]{revtex4}
\usepackage{graphicx}% Include figure files
\usepackage{amscd,amsmath,amsthm,amsfonts,amssymb}
\usepackage{ytableau}
\usepackage{extarrows}
\def\[{\begin{equation}}
\def\]{\end{equation}}
\newtheorem{lem}{Lemma}
\newtheorem{thm}{Theorem}
\begin{document}
\title{Concentric-ring patterns of higher-order lumps in the Kadomtsev--Petviashvili I equation}
\author{%%%% Author details
Bo Yang$^{1}$, Jianke Yang$^{2}$}
%%%%%%%%% Insert author address here
\address{$^{1}$ School of Mathematics and Statistics, Ningbo University, Ningbo 315211, China\\
$^{2}$ Department of Mathematics and Statistics, University of Vermont, Burlington, VT 05401, U.S.A}
\begin{abstract}
Large-time patterns of general higher-order lump solutions in the KP-I equation are investigated. It is shown that when the index vector of the general lump solution is a sequence of consecutive odd integers starting from one, the large-time pattern in the spatial $(x, y)$ plane generically would comprise fundamental lumps uniformly distributed on concentric rings. For other index vectors, the large-time pattern would comprise fundamental lumps in the outer region as described analytically by the nonzero-root structure of the associated Wronskian-Hermit polynomial, together with possible fundamental lumps in the inner region that are uniformly distributed on concentric rings generically. Leading-order predictions of fundamental lumps in these solution patterns are also derived. The predicted patterns at large times are compared to true solutions, and good agreement is observed.
\end{abstract}
\maketitle

\section{Introduction}
The Kadomtsev--Petviashvili (KP) equation was derived as a two-dimensional generalization of the Korteweg-de Vries equation for the evolution of weakly nonlinear plasma waves and shallow water waves \cite{KP1970,Ablowitz1979}. In the water wave context, this equation reads \cite{Ablowitz1979}
\[
\left(2f_{t}+3ff_x+(\frac{1}{3}-T)f_{xxx}\right)_x+f_{yy}=0,
\]
where the spatial coordinate $x$ is relative to a certain moving frame, $y$ is the transverse spatial coordinate, $t$ is time, $f(x,y,t)$ represents the water surface elevation, and $T$ is a dimensionless surface tension parameter. If $T>1/3$, which corresponds to thin sheets of water, this equation is called KP-I. In this case, rescaling variables by
\[
y=\frac{\hat{y}}{\sqrt{3(T-\frac{1}{3}})}, \quad t=-\frac{2\hat{t}}{T-\frac{1}{3}}, \quad f=-2(T-\frac{1}{3})u
\]
and dropping the hats, this equation is converted to the standard form
\[\label{KPI}
\left( u_t+ 6 u u_x +u_{xxx} \right)_x - 3u_{yy} =0.
\]
Note that the KP-I equation also arises in other branches of physics, such as nonlinear optics \cite{Pelinovsky1995} and Bose-Einstein condensates \cite{BarashenkovBEC,Tsuchiya_BEC2008}.

The KP-I equation (\ref{KPI}) is solvable by the inverse scattering transform \cite{Zakharov_book, Ablowitz_book}. It admits stable fundamental lump solutions that are bounded rational functions decaying in all spatial directions \cite{Petviashvili1976,Manakov1977,Ablowitz_Satsuma1979}. These lumps are the counterparts of solitons in the Korteweg-de Vries equation. In the water wave context, these lumps physically correspond to dips on the water surface due to the negative sign in the $f$ scaling above. The KP-I equation also admits a broad class of rational solutions that describe the interactions of these lumps. If individual lumps have distinct asymptotic velocities, they would pass through each other without change in velocities or phases \cite{Manakov1977,Ablowitz_Satsuma1979}. But if they have the same asymptotic velocities, they would undergo novel anomalous scattering, where the lumps would separate from each other in new spatial directions that are very different from their original incoming directions \cite{Peli93b,Ablowitz97,Ablowitz2000}. In this article, we are concerned with this latter type of solutions, which we will call higher-order lumps (they are also called multi-pole lumps in the literature \cite{Ablowitz97,Ablowitz2000}).

Explicit expressions of higher-order lumps have been derived by a wide variety of methods before \cite{Peli93a,Peli93b,Ablowitz97,Ablowitz2000,Dubard2010,Dubard2013,Chen2016,ClarksonDowie2017,Gaillard2018,
Chang2018,Guo2022,YangYangKPI,Gaillard2022}. In addition, large-time patterns for special classes of these higher-order lumps have also been investigated \cite{Chen2016,Chang2018,YangYangKPI,Chakravarty2022,Chakravarty2023}. Of particular interest to us are the analytical results in \cite{YangYangKPI}, where we showed that for a certain class of higher-order lumps, when the index vector is a sequence of consecutive odd integers starting from one, the solution pattern at large time would comprise fundamental lumps arranged in a triangular shape, which is described analytically by the root structure of an Yablonskii-Vorob'ev polynomial. For other index vectors, the solution pattern at large time would comprise fundamental lumps arranged in a nontriangular shape in the outer region, which is described analytically by the nonzero-root structure of the associated Wronskian-Hermit polynomial, together with possible fundamental lumps arranged in a triangular shape in the inner region, which is described analytically by the root structure of an Yablonskii-Vorob'ev polynomial (this inner region would appear when the associated Wronskian-Hermit polynomial has a zero root).

In this paper, we investigate large-time patterns of \emph{general} higher-order lump solutions in the KP-I equation. We show that when the index vector of the general lump solution is a sequence of consecutive odd integers starting from one, the large-time pattern \emph{generically} would comprise fundamental lumps uniformly distributed on concentric rings (in other words, these fundamental lumps would form regular polygons with the same center). This concentric-ring pattern strongly contrasts the triangular pattern of special higher-order lumps considered in \cite{YangYangKPI} and is quite surprising. We also show that on these concentric rings, the fundamental lumps separate from each other in proportion to $|t|^{m/(2m+1)}$, where $m$ is a positive integer that takes on different values on different rings. For other index vectors, we show that the large-time pattern of a general higher-order lump solution would comprise fundamental lumps in the outer region as described analytically by the nonzero-root structure of the associated Wronskian-Hermit polynomial, together with possible fundamental lumps in the inner region that are uniformly distributed on concentric rings generically. Leading-order predictions of fundamental lumps in these solution patterns are also derived. Our predicted patterns at large times are compared to true solutions, and good agreement is observed.

This paper is structured as follows. In Sec.~\ref{sec2}, we present explicit algebraic expressions of higher-order lumps, which is the starting point of our analysis. In Sec.~\ref{sec3}, we present our analytical predictions of generic lump patterns at large time when the index vector is a sequence of consecutive odd integers starting from one, and verify these predictions quantitatively by two examples. In Sec.~\ref{secProof}, we prove our analytical predictions of lump patterns given in Sec~\ref{sec3}. In Sec.~\ref{sec5}, we present our analytical predictions of generic lump patterns at large time for other index vectors, and verify these predictions quantitatively by two examples. In Sec.~\ref{secProof2}, we prove our analytical predictions of lump patterns stated in Sec~\ref{sec5}. Sec.~\ref{sec7} concludes the paper with discussions on nongeneric lump patterns.

\section{Preliminaries} \label{sec2}
Explicit expressions of higher-order lumps in KP-I have been derived in \cite{YangYangKPI} by the KP hierarchy reduction method. Since the KP-I equation (\ref{KPI}) is invariant under the Galilean transformation \cite{Weiss1985,Chen2016}
\[ \label{Galilean}
(x,y,t) \to (x+2ky+12k^2t, \, y+12kt, \, t)
\]
for an arbitrary real constant $k$, and also invariant under variable rescalings
\[ \label{scaling}
(x, y, t, u)\to (\alpha x, \alpha ^2y, \alpha^3t, \alpha^{-2}u)
\]
for an arbitrary nonzero real constant $\alpha$, we can normalize the spectral parameter $p$ in those higher-order lumps to be unity \cite{YangYangKPI}. In that case, introducing elementary Schur polynomials $S_k(\mbox{\boldmath $x$})$ with $ \emph{\textbf{x}}=\left( x_{1}, x_{2}, \ldots \right)$, which  are defined by the generating function
\begin{equation}\label{Elemgenefunc}
\sum_{k=0}^{\infty}S_k(\mbox{\boldmath $x$}) \epsilon^k
=\exp\left(\sum_{n=1}^{\infty}x_n \epsilon^n\right),
\end{equation}
and defining $S_k(\mbox{\boldmath $x$})\equiv 0$ when $k<0$, explicit expressions of those higher-order lumps are reproduced in the following lemma.

\begin{lem} \label{Lemma1}
General $N$-th order lumps of the KP-I equation (\ref{KPI}) are
\[ \label{Schpolysolu}
u_{\Lambda} (x,y,t)=2 \partial_{x}^2 \ln \sigma,
\]
where
\[\label{Blockmatrix}
\sigma(x,y,t)=\det_{1 \leq i,j \leq N}\left(m_{ij}\right),
\]
\[ \label{Schmatrimnij}
m_{i, j}=\sum_{\nu=0}^{\min(n_i,n_j)} 4^{-\nu} \hspace{0.06cm} S_{n_i-\nu}\left(\textbf{\emph{x}}^{+} +\nu \textbf{\emph{s}}+\textbf{\emph{a}}_{i} \right)  \hspace{0.06cm} S_{n_j-\nu}\left((\textbf{\emph{x}}^{+})^* + \nu \textbf{\emph{s}}^*+ \textbf{\emph{a}}_{j}^*\right),
\]
$N$ is an arbitrary positive integer, $\Lambda \equiv (n_{1}, n_{2}, \cdots n_N)$ is an index vector of arbitrary positive integers, the asterisk `*' represents complex conjugation, the vector $\textbf{\emph{x}}^{+}=\left( x_{1}^{+}, x_{2}^{+},\cdots \right)$ is defined by
\[
x_{k}^{+}= \frac{1}{k!} x + \frac{2^k}{k!} \textrm{i} y +  \frac{3^k}{k!} (-4) t,  \label{defxrp}
\]
the vector $\textbf{\emph{s}}=(s_1, s_2, \cdots)$ is defined through the expansion
\begin{eqnarray} \label{sexpand}
\sum_{j=1}^{\infty} s_{j}\alpha^{j}=\ln \left[\frac{2}{\alpha}  \tanh \left(\frac{\alpha}{2}\right)\right],
\end{eqnarray}
internal parameters $\textbf{\emph{a}}_{i}$ are
\[\label{multiparas}
\textbf{\emph{a}}_{i}=\left( a_{i,1}, a_{i,2}, \cdots , a_{i,n_i}\right),
\]
and $a_{i, j} \hspace{0.05cm} (1\le i\le  N,  1\le j\le n_i)$ are free complex constants.
\end{lem}

Without loss of generality, we require $(n_{1}, n_{2}, \cdots n_N)$ to be distinct integers \cite{YangYangKPI}.

The fundamental lump can be derived by taking $N=1$ and $n_1=1$ in the above lemma. Through a shift of the $(x,y)$ axes, we can normalize $a_{1,1}=1$. Then the formula for this fundamental lump is
\[ \label{defu1}
u_1(x,y,t)=2 \partial_{x}^2 \ln \left(\left(x-12t  \right)^2+4y^2+ \frac{1}{4}\right)=
\frac{16[1-4(x-12t)^2+16y^2]}{[1+4(x-12t)^2+16y^2]^2}.
\]
Its graph is a single main hump centered at $(x, y)=(12t, 0)$ with peak amplitude $16$.

\section{Concentric rings of lumps at large times for $\Lambda=(1, 3, \dots, 2N-1)$} \label{sec3}
In our earlier paper \cite{YangYangKPI}, we have shown that in the special case of parameter regimes
where $a_{i,j}$ in Eq.~(\ref{multiparas}) is independent of the $i$ index, i.e., $a_{1, j}=a_{2,j}=\dots=a_{n_N, j}$, then
patterns of higher-order lumps in Lemma 1 at large times would comprise fundamental lumps arranged in nontriangular shapes in the outer region, whose locations are determined analytically by the nonzero-root structure of the underlying Wronskian-Hermit polynomial, together with fundamental lumps arranged in triangular shapes in the inner region, whose locations are determined analytically by the root structure of the underlying Yablonskii-Vorob'ev polynomial. Outer nontriangular lumps are present when $\Lambda\ne (1, 3, \dots, 2N-1)$ and absent when $\Lambda=(1, 3, \dots, 2N-1)$, while inner triangular lumps are present if the underlying Wronskian-Hermit polynomial admits a zero root and absent if that polynomial has no zero root.

Our interest in this paper is to identify new patterns of higher-order lumps at large times. We will show that in the more general parameter regime where $a_{i,1}$ is dependent on the $i$ index, i.e., $\{a_{i,1}, 1\le i\le N\}$ are not all the same, concentric rings of fundamental lumps would generically appear. These concentric rings of fundamental lumps mean certain numbers of fundamental lumps that are uniformly distributed on concentric circles of the $(x, y)$ plane (when the $y$ direction is properly stretched). In other words, these fundamental lumps are located at vertices of multiple regular polygons with the same center. When $\Lambda=(1, 3, \dots, 2N-1)$, these concentric rings of fundamental lumps are the only solution patterns in the spatial $(x, y)$ plane, while when $\Lambda\ne (1, 3, \dots, 2N-1)$, we may get these concentric rings of lumps in the inner region, together with certain numbers of fundamental lumps in the outer region. Since the $\Lambda=(1, 3, \dots, 2N-1)$ case is simpler, we will treat it first.

\subsection{Notations and our assumption}
Before describing our results, we introduce some notations.

First, we define $[a]$ as the largest integer less than or equal to $a$.

We also define the following matrices
\begin{eqnarray}
&& \mathbf{D}= \mbox{diag}\left(a_{1,1}, a_{2,1}, a_{3,1},\cdots, a_{N,1} \right), \label{defD} \\
&& \mathbf{F}=\left(
\begin{array}{ccccc}
1 &  0 & 0 & \cdots & 0 \\
\frac{1}{1!} & 1 & 0 & \cdots & 0 \\
\frac{1}{2!} & \frac{1}{1!} & 1 & \cdots & 0 \\
\vdots &  \vdots & \ddots & \ddots & \vdots \\
\frac{1}{\left(N-1\right)!} & \frac{1}{\left(N-2\right)!} & \frac{1}{\left(N-3\right)!}  & \cdots & 1
\end{array}
\right)_{N\times N}, \label{defF}  \\
&& \mathbf{G} = \mbox{Mat}_{1\leq i,j\leq N}\left(g_{i,j}\right)\equiv\mathbf{F}^{-1} \mathbf{D} \mathbf{F}. \label{defG}
\end{eqnarray}
Notice that both $\mathbf{F}$ and $\mathbf{G}$ are lower triangular matrices.

In addition, we introduce the following $[N/2]$ minors of the $\mathbf{G}$ matrix,
\[\label{defMr0}
M_r\equiv \det_{N+1-r\le i\le N, \ 1\le j\le r}\mathbf{G}, \quad r=1, 2, \cdots, [N/2].
\]
These minors are determinants of square submatrices in the lower left corner of matrix $\mathbf{G}$.

With the above notations, our only assumption in this section is the following.

\begin{quote}
\textbf{Assumption 1.} \emph{We assume that}
\[ \label{defMr}
M_r\ne 0, \quad r=1, 2, \dots, [N/2]-1; \quad \frac{M_{[N/2]}}{M_{[N/2]-1}} \ne
\left\{\begin{array}{ll}
0, &  \mbox{when $N$ is odd}, \\
-\frac{4}{3}, &   \mbox{when $N$ is even}. \end{array}\right.
\]
\end{quote}

This assumption holds for generic values of $\left(a_{1,1}, a_{2,1}, a_{3,1},\cdots, a_{N,1} \right)$.

The consequence of this assumption is that, the $[N/2]\times [N/2]$ submatrix in the lower left corner of $\mathbf{G}$ would admit the following factorization
\[
\mathbf{G}_{N+1-[N/2]\le i\le N, \ 1\le j\le [N/2]}= \mathbf{A}\mathbf{B},   \label{FacAB}
\]
where
\[
\mathbf{A}=\left(
\begin{array}{ccccc}
 1 & \alpha _{1,2} & \cdots & \alpha _{1,[N/2]}  \\
 0 & 1 & \cdots & \alpha _{2,[N/2]}  \\
 \vdots & \vdots & \ddots & \vdots \\
 0 & 0 & 0 & 1  \\
\end{array}
\right),\quad \mathbf{B} =\left(
\begin{array}{ccccc}
 0 & 0 & \cdots & \beta _{[N/2],[N/2]}  \\
 \vdots & \vdots & .\cdot^{\cdot} & \vdots \\
 0 & \beta _{2,2} & \cdots & \beta _{2,[N/2]}  \\
 \beta _{1,1} & \beta _{1,2} & \cdots & \beta _{1,[N/2]} \\
\end{array}
\right),
\]
and $\alpha_{i,j}$, $\beta_{i,j}$ are complex constants. In particular,
\[
\beta _{1,1}=M_1, \quad \beta_{r,r}=\frac{M_{r}}{M_{r-1}}, \quad 1<r\le [N/2].
\]
Under Assumption 1, $\beta_{r,r}\ne 0$ for $1\le r\le [N/2]-1$, and $\beta_{[N/2], [N/2]}\ne 0$ when $N$ is odd and
$\beta_{[N/2], [N/2]}+4/3\ne 0$ when $N$ is even. The factorization (\ref{FacAB}) is similar to Gauss elimination, but starting from the bottom row of the matrix up instead of from the top row down. This unconventional matrix factorization turns out to be important for our current problem.

We note that in our previous work \cite{YangYangKPI}, $\{a_{i,1}, 1\le i\le N\}$ are all the same. In that case, $\mathbf{G}$ is a diagonal matrix and thus $M_1=0$, violating the above assumption. Thus, this assumption means that we are now dealing with a new parameter regime different from \cite{YangYangKPI}.

\subsection{Main results} \label{secMainResult}
Under Assumption 1, our main results on lump patterns for $\Lambda=(1, 3, \dots, 2N-1)$ at large times are given in the following theorem.

\begin{thm} \label{Theorem1}
If $\Lambda=(1, 3, \dots, 2N-1)$ with $N>1$ and Assumption 1 holds, then when $|t|\gg 1$, the higher-order lump solution $u_{\Lambda} (x,y,t)$ in Lemma 1 would split into $N(N+1)/2$ fundamental lumps on the $(x, y)$ plane. These fundamental lumps are asymptotically located uniformly on $[N/2]$ concentric rings centered at $(x, y)=(12t, 0)$, with one of them also located in the $O(1)$ neighrborhood of the ring center when $N$ is odd. The $r$-th ring (counting from outside with $1\le r\le [N/2]$) contains $2N-1-4(r-1)$ fundamental lumps $u_{1}(x-x_{0}, \hspace{0.04cm}  y-y_{0}, t)$, where $u_1(x, y, t)$ is given in Eq.~(\ref{defu1}), and its $(x_0, y_0)$ positions (relative to the ring center $(x, y)=(12t, 0)$) are given by the equation
\[\label{x0t0r}
x_0+2{\rm{i}}y_0= z_0 \hspace{0.05cm} (-12t)^{\frac{N-1-2(r-1)}{2N-1-4(r-1)}}\left(1+O\left(|t|^{-\frac{1}{2N-1-4(r-1)}}\right)\right).
\]
Here, $z_0$ is every one of the $\left(2N-1-4(r-1)\right)$-th roots of $-\beta_{r,r}\left(2N-1-4(r-1)\right)!!\left(2N-3-4(r-1)\right)!!$, except in the case of even $N$ and on its $[N/2]$-th (inner-most) ring, in which case $z_0$ is every one of the cubic roots of $-(\beta_{[N/2],[N/2]}+4/3)3!!1!!$. Written mathematically, we have the following solution asymptotics
\[
u_{\Lambda} (x,y,t)=u_1(x-x_0, y-y_0)+O\left(|t|^{-\frac{1}{2N-1-4(r-1)}}\right), \quad |t|\gg 1,
\]
where $1\le r\le [N/2]$, and $(x_0, y_0)$ is as given above for each of the stated roots $z_0$.
\end{thm}

The proof of this theorem will be given in the next section. Note that since $\beta_{r,r}\ne 0$ for $1\le r\le [N/2]-1$ and $\beta_{[N/2], [N/2]}+4/3\ne 0$ under Assumption 1, all roots $z_0$ mentioned in this theorem are nonzero.

This theorem indicates when $\Lambda=(1, 3, \dots, 2N-1)$, the solution pattern of higher-order lumps at large time is a set of $[N/2]$ concentric rings of fundamental lumps in the generic case where Assumption 1 holds. This strongly contrasts the case when internal parameters $\{a_{i,j}\}$ are independent of the $i$ index, in which case the solution pattern is a triangular pattern of fundamental lumps as we have reported earlier in \cite{YangYangKPI}.

We should point out that these rings of lumps are not circular on the $(x, y)$ plane. Indeed, we can see from Eq.~(\ref{x0t0r}) that the locations $(x_0, y_0)$ of these lumps are not on a circle. Rather, $(x_0, 2y_0)$ of those lumps are on a circle. Thus, these rings are ellipses which are longer along the $x$ direction than the $y$ direction by a factor of 2. We call them rings rather than ellipses for simplicity. In later graphs of this paper (see Figs.~\ref{f:fig1}-\ref{f:fig2} for instance), the lumps seem to be on a circle at large times, but that is only because our $x$ interval is twice as long as the $y$ interval there, i.e., we have stretched the $y$ direction by a factor of two.

Notice that the leading-order positions of these fundamental lumps on the rings, as given in Eq.~(\ref{x0t0r}), are determined only by the first elements $\{a_{i,1}\}$ of the internal parameter vectors $\textbf{\emph{a}}_{i}$. This is not surprising, since any Schur polynomial $S_{k}\left(\textbf{\emph{x}}^{+} +\nu \textbf{\emph{s}}+\textbf{\emph{a}}_{i} \right)$ is affected most by its first element $x_1^{+}+a_{i,1}$.

The above theorem shows that, of these $[N/2]$ concentric rings, the outer-most ring (with $r=1$) contains $2N-1$ fundamental lumps $u_{1}(x-x_{0}, \hspace{0.04cm}  y-y_{0}, t)$, where $u_1(x, y, t)$ is given in Eq.~(\ref{defu1}), and its $(x_0, y_0)$ positions (relative to the ring center $(x, y)=(12t, 0)$) are given by
\[\label{x0t01}
x_0+2{\rm{i}}y_0= z_0 \hspace{0.05cm} (-12t)^{\frac{N-1}{2N-1}}\left(1+O\left(|t|^{-\frac{1}{2N-1}}\right)\right),
\]
with $z_0$ being every one of the $(2N-1)$-th roots of $-\beta_{1,1}(2N-1)!!(2N-3)!!$. The second outer-most ring (with $r=2$) contains $2N-5$ fundamental lumps $u_{1}(x-x_{0}, \hspace{0.04cm}  y-y_{0}, t)$, whose $(x_0, y_0)$ positions relative to the ring center $(x, y)=(12t, 0)$ are given by
\[\label{x0t02}
x_0+2{\rm{i}}y_0= z_0 \hspace{0.05cm} (-12t)^{\frac{N-3}{2N-5}}\left(1+O\left(|t|^{-\frac{1}{2N-5}}\right)\right),
\]
with $z_0$ being every one of the $(2N-5)$-th roots of $-\beta_{2,2}(2N-5)!!(2N-7)!!$, and so on. For each next ring inward, the number of fundamental lumps on it decreases by 4. When $N$ is odd, the inner-most ring (with $r=[N/2]$) contains 5 fundamental lumps, whose $(x_0, y_0)$ positions are given by
\[\label{x0t0r2}
x_0+2{\rm{i}}y_0= z_0 \hspace{0.05cm} (-12t)^{\frac{2}{5}}\left(1+O\left(|t|^{-\frac{1}{5}}\right)\right),
\]
with $z_0$ being every one of the quintic roots of $-\beta_{[N/2],[N/2]}5!!3!!$. In this case, there is
an additional fundamental lump in the $O(1)$ neighborhood of the ring center. When $N$ is even, the inner-most ring (with $r=[N/2]$) contains 3 fundamental lumps, whose $(x_0, y_0)$ positions are given by
\[\label{x0t0r1}
x_0+2{\rm{i}}y_0= z_0 \hspace{0.05cm} (-12t)^{\frac{1}{3}}\left(1+O\left(|t|^{-\frac{1}{3}}\right)\right),
\]
with $z_0$ being every one of the cubic roots of $-(\beta_{[N/2],[N/2]}+4/3)3!!1!!$. In this case, there is no additional fundamental lump at the ring center.

Theorem~\ref{Theorem1} also shows that, when $N$ is odd, the locations of fundamental lumps on the concentric rings at large times $t=\pm t_0$ are the same, because $N-1-2(r-1)$ is even in this case and thus $(-12t)^{\frac{N-1-2(r-1)}{2N-1-4(r-1)}}$ is the same for $t=\pm t_0$. But when $N$ is even, the locations of fundamental lumps on the concentric rings at large time $t=t_0$ would be antisymmetric to those at large time $t=-t_0$, since $N-1-2(r-1)$ is odd now and thus $(-12t)^{\frac{N-1-2(r-1)}{2N-1-4(r-1)}}$ is opposite of each other for $t=\pm t_0$. This means that when $N$ is odd, the locations of fundamental lumps on concentric rings at large times $\pm t_0$ would be the same, i.e., there would be no change of lump positions on the rings as time increases from large negative to large positive. But when $N$ is even, the locations of fundamental lumps on concentric rings at large times $\pm t_0$ would be antisymmetric to each other, i.e., a change of lump positions on the rings would occur as time increases from large negative to large positive. This difference regarding lump positions on the rings for odd and even $N$ will be seen in Figs.~\ref{f:fig1}-\ref{f:fig2} and \ref{f:fig4}-\ref{f:fig5} in the next subsection.

Eq.~(\ref{x0t0r}) of the above theorem reveals that at large time $|t|$, fundamental lumps on the $r$-th ring separate from each other in proportion to $|t|^{\frac{N-1-2(r-1)}{2N-1-4(r-1)}}$. Thus, by choosing $N$ and $r$ properly, we can get separation rates of $|t|^{\frac{m}{2m+1}}$ for any positive integer $m$, such as $|t|^{1/3}$, $|t|^{2/5}$, $|t|^{3/7}$, and so on. This contrasts the results in \cite{YangYangKPI} for special internal-parameter values, where the separation rate of fundamental lumps is only $|t|^{1/3}$. In \cite{Ablowitz2000}, it was reported that at large time, fundamental lumps in the higher-order lump complex separate from each other in proportion to $|t|^q$, where $\frac{1}{3}\le q\le \frac{1}{2}$. Our separation rates from the above theorem are consistent with this $q$ range, but they are more specific with the form of $q$ as $m/(2m+1)$, not any real number between $1/3$ and $1/2$.

Regarding the fundamental lump in the $O(1)$ neighborhood of the ring center for odd $N$, Theorem~\ref{Theorem1} did not provide an asymptotic prediction for its position. Actually, large-time prediction of its position can be made by slightly modifying the calculations leading to Eq.~(\ref{x0t0r}). This modification is necessary since the position of the center lump is $O(1)$ from the ring center, unlike lumps on the rings whose positions are $O(|t|^q)$ away from the ring center with $1/3\le q< 1/2$. This difference means that slightly different asymptotic calculations are in order for the position of the center lump. With a little algebra, we can show that this prediction can be obtained from a slightly modified submatrix of $\mathbf{G}$. Specifically, we take the bottom left $([N/2]+1)\times ([N/2]+1)$ submatrix of $\mathbf{G}$ in Eq.~(\ref{defG}), and increase its two matrix elements adjacent to its top-right corner (i.e., its $(1, [N/2])$ and $(2, [N/2]+1)$ elements) by $4/3$ and call this new matrix $\widetilde{\mathbf{G}}$. Then, we perform the $\widetilde{\mathbf{A}}\widetilde{\mathbf{B}}$ factorization to this new matrix $\widetilde{\mathbf{G}}$ similar to Eq.~(\ref{FacAB}). Then, the leading-order position $(x_0, y_0)$ of the center lump would be predicted by $x_0+2{\rm{i}}y_0=-\tilde{\beta}_{[N/2]+1, [N/2]+1}$, where $\tilde{\beta}_{[N/2]+1, [N/2]+1}$ is from the $\widetilde{\mathbf{B}}$ matrix of that factorization, and the error of this $(x_0, y_0)$ prediction is $O(|t|^{-1}$).
This prediction of the center lump was not written into Theorem~\ref{Theorem1} because we do not want it to distract the reader's attention from the main focus of the paper, which is the concentric rings of lumps.

\subsection{Numerical verifications of Theorem 1}
Now, we numerically verify Theorem~\ref{Theorem1} on two examples.

\begin{figure}[htb]
\begin{center}
\vspace{8.0cm}
\hspace{-2.0cm}
\includegraphics[scale=0.25, bb=300 0 400 360]{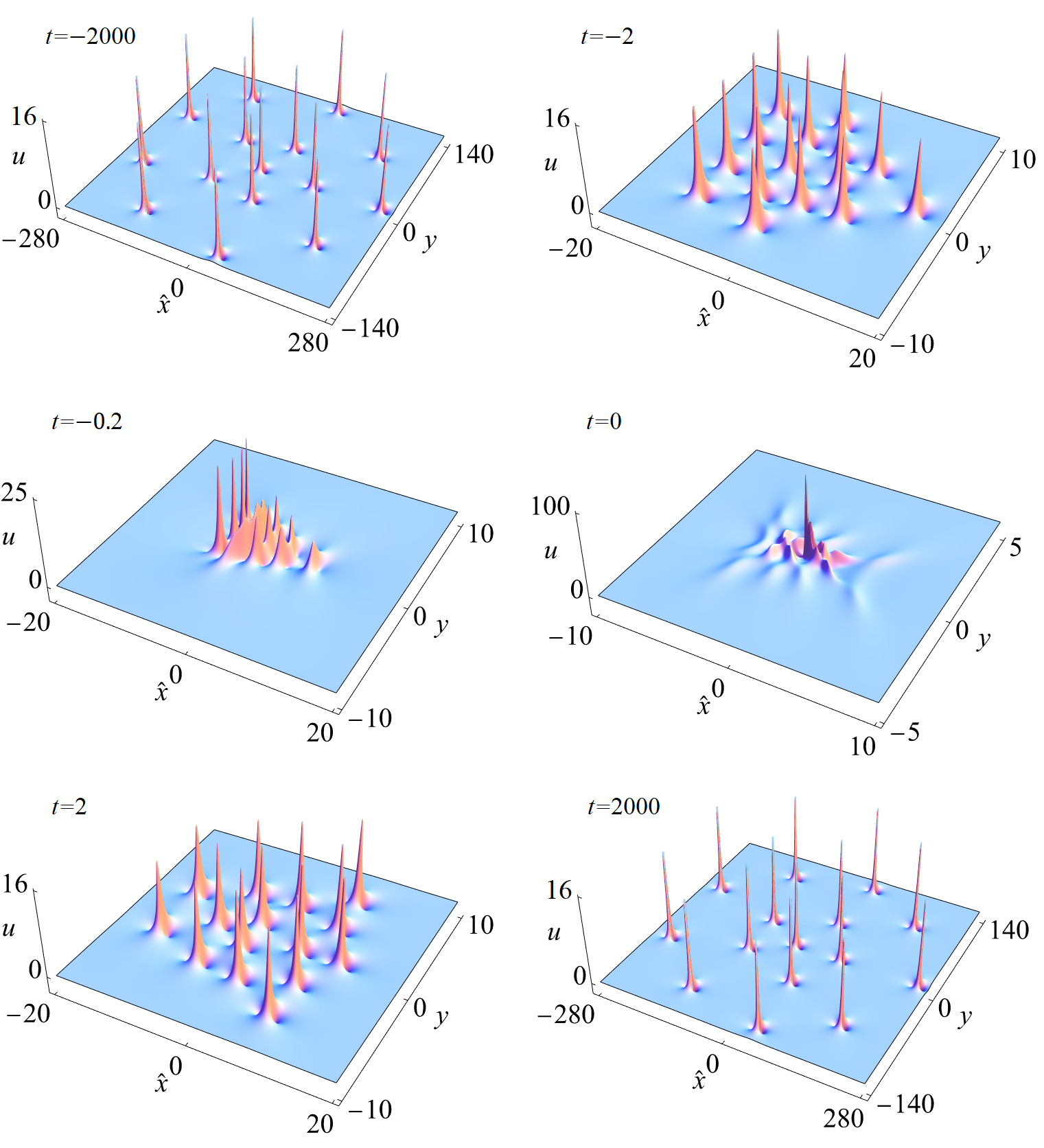}\hspace{2.0cm}
\caption{The true solution $u_{\Lambda} (x,y,t)$ with $\Lambda =(1, 3, 5, 7, 9)$ at time values of $t=-2000, -2, -0.2, 0, 2$ and 2000. Internal parameters are given in Eq.~(\ref{para1}), with the other elements of parameter vectors as zero. The axis $\hat{x}=x-12t$ is the moving $x$ coordinate. }  \label{f:fig1}
\end{center}
\end{figure}

\textbf{Example 1.} In our first example, we take $N=5$; so $\Lambda =(1, 3, 5, 7, 9)$. Internal parameters are taken as
\[ \label{para1}
(a_{1,1}, a_{2,1}, a_{3,1}, a_{4,1}, a_{5,1})=(0, 1, 1, 1, -1),
\]
with the other elements of parameter vectors $\textbf{\emph{a}}_i \ (1\le i\le 5)$ taken as zero.
The true solution from Lemma~1 at six time values of $t=-2000, -2, -0.2, 0, 2 $ and 2000 is plotted in Fig.~\ref{f:fig1}.
It is seen that at $t=-2000$, the solution splits into two concentric rings with 9 and 5 fundamental lumps evenly distributed on them respectively, plus another fundamental lump located at the ring center. As time increases to $-2$, these 15 fundamental lumps get close to each other, and the shape formed by them has changed as well. As time increases further to $-0.2$ and 0, these 15 fundamental lumps merge with each other and form some high spikes. However, as time further increases to $2$, the merged solution splits up into 15 fundamental lumps again in a quasi-trapezoid shape. When time continues to increase to 2000, these 15 fundamental lumps evolve into two concentric rings with 9 and 5 fundamental lumps on them, plus another fundamental lump located at the ring center, similar to the pattern at $t=-2000$. In particular, the relative positions of fundamental lumps on the two concentric rings at $t=\pm 2000$ are roughly the same.

\begin{figure}[htb]
\begin{center}
\includegraphics[scale=0.25, bb=0 0 1150 490]{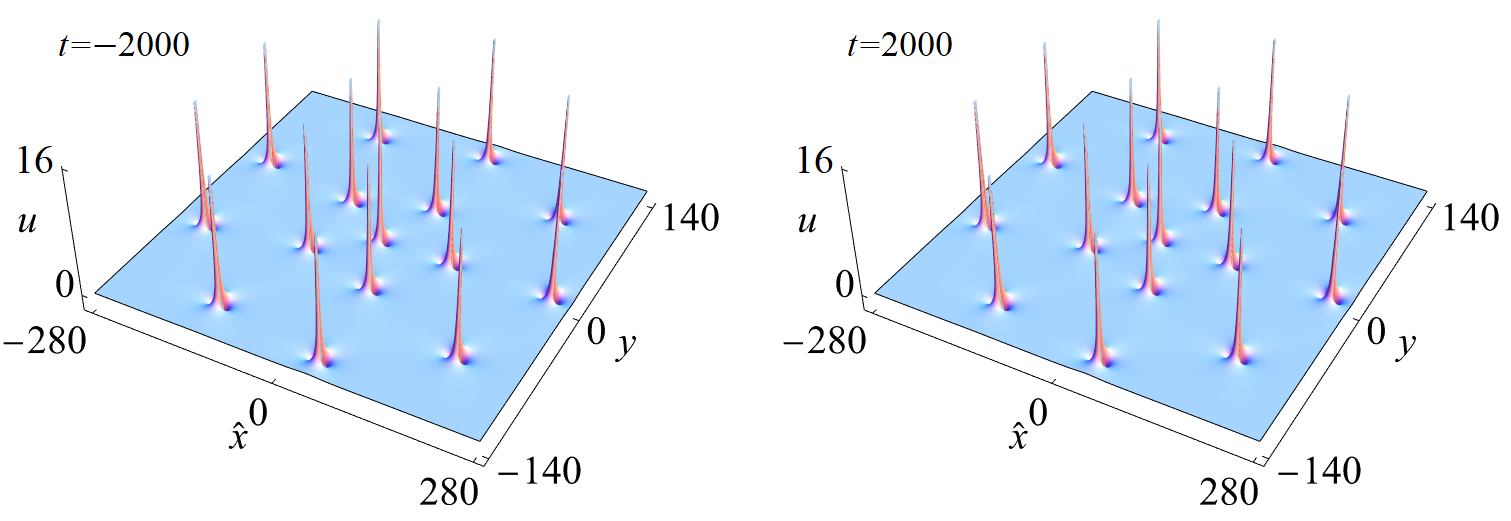}
\caption{Leading-order predictions of lump patterns from Theorem~\ref{Theorem1} for the solution of Fig.~\ref{f:fig1} at large times of $t=\pm 2000$.  }  \label{f:fig2}
\end{center}
\end{figure}

Now, we use Theorem~\ref{Theorem1} to analytically predict the solution of Fig.~\ref{f:fig1} at large times of $t=\pm 2000$. That theorem predicts that the solution of Fig.~\ref{f:fig1} at large times would split into two concentric rings with 9 and 5 fundamental lumps on them respectively, together with a fundamental lump near the ring center. To determine analytical predictions of lump positions on these two rings, we notice from the parameter choices (\ref{para1}) that $\mathbf{D}= \mbox{diag}(0, 1, 1, 1, -1)$. Then, the $2\times 2$ submatrix at the lower left corner of matrix $\mathbf{G}$ from Eq.~(\ref{defG}) and its factored $\mathbf{B}$ matrix from Eq.~(\ref{FacAB}) are
\[
\mathbf{G}_{4\le i\le 5, \ 1\le j\le 2}=
\left(\begin{array}{cc} \frac{1}{6} & 0  \\
 -\frac{1}{8} & -\frac{1}{3} \end{array}\right), \quad \mathbf{B}=\left(\begin{array}{cc} 0 & -\frac{4}{9}  \\
 -\frac{1}{8} & -\frac{1}{3} \end{array}\right).
\]
This shows that $\beta_{1,1}=-1/8$ and $\beta_{2,2}=-4/9$. Notice that the above submatrix of $\mathbf{G}$ satisfies our Assumption 1. Using these values, we can obtain leading-order predictions of lump positions on these two rings from Eq.~(\ref{x0t0r}) with $N=5$ and $r=1, 2$. These predicted solutions at large times of $t=\pm 2000$ are plotted in Fig.~\ref{f:fig2} (the center lump whose position is predicted by the last paragraph of the previous subsection is also shown for completeness). Comparing these predictions with true solutions at $t=\pm 2000$ in Fig.~\ref{f:fig1}, we can see that the predictions agree with true solutions very well.

\begin{figure}[htb]
\begin{center}
\includegraphics[scale=0.3, bb=100 0 1150 490]{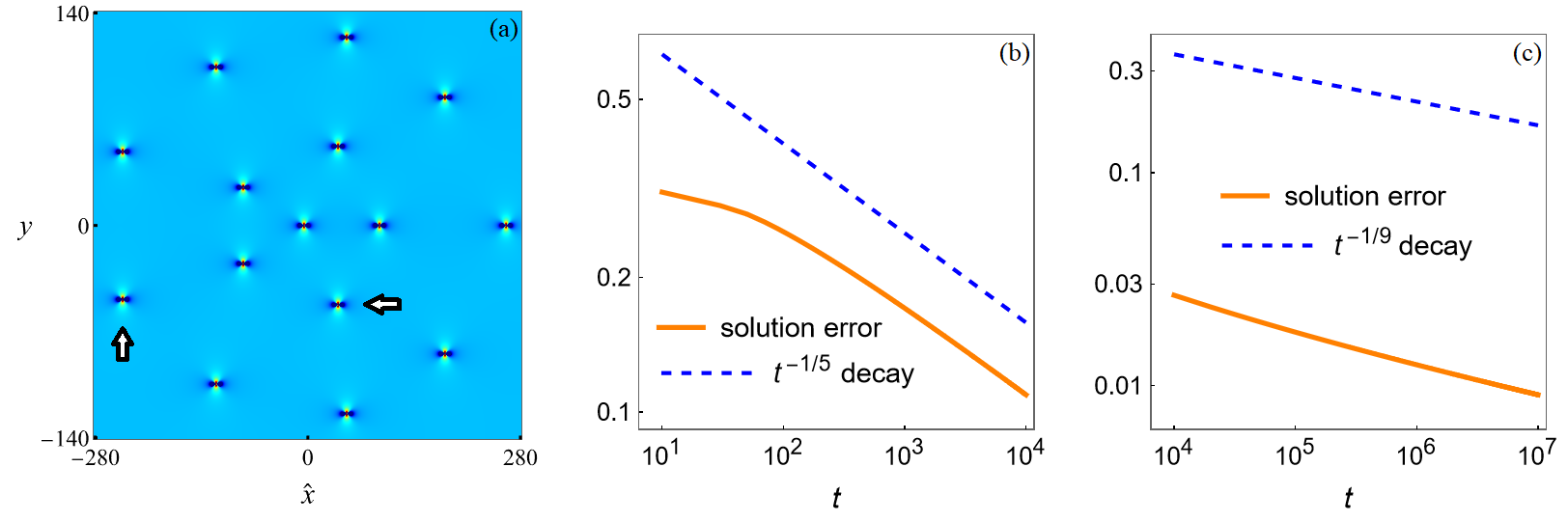}
\caption{Verification of the decay rate of relative error in leading-order predictions of fundamental lumps' positions in Eq.~(\ref{x0t0r}) of Theorem~\ref{Theorem1} for the example of Fig.~\ref{f:fig1} with $\Lambda=(1, 3, 5, 7, 9)$. (a) Density plot of the true higher-order lump solution in Fig.~\ref{f:fig1} at $t=2000$. (b) Relative error versus time $t$ for the location of the lump on the inner ring marked by a horizontal arrow in (a) (the predicted $|t|^{-1/5}$ decay is plotted as a dashed line for comparison). (c) Relative error versus time $t$ for the location of the lump on the outer ring marked by a vertical arrow in (a) (the predicted $|t|^{-1/9}$ decay is plotted as a dashed line for comparison).}  \label{f:fig3}
\end{center}
\end{figure}

Next, we quantitatively compare predicted and true solutions at various time values in order to verify the decay rate of relative errors on fundamental lumps' positions in Eq.~(\ref{x0t0r}). For this purpose, the density plot of the true higher-order lump solution in Fig.~\ref{f:fig1} at $t=2000$ is displayed in Fig.~\ref{f:fig3}(a). We then pick a fundamental lump on the inner ring (marked by a horizontal white arrow), and a fundamental lump on the outer ring (marked by a vertical white arrow). For each fundamental lump, we numerically determine at each large time $t$ the relative error of prediction for its position, which is defined as $\sqrt{(x_{0, true}-x_0)^2+(y_{0, true}-y_0)^2}/\sqrt{(x_{0, true})^2+(y_{0, true})^2}$, where $(x_{0, true}, y_{0, true})$ is the true location of the lump (relative to the ring center $(x,y) = (12t,0)$), and $(x_0, y_0)$ is its leading-order prediction from Eq.~(\ref{x0t0r}). The graphs of this relative error versus time $t$ for these two fundamental lumps are plotted in Fig.~\ref{f:fig3}(b, c), respectively. For the lump on the inner ring (with $N=5$ and $r=2$), the predicted relative error from Eq.~(\ref{x0t0r}) is $O(|t|^{-1/5})$, while for the lump on the outer ring (with $N=5$ and $r=1$), the predicted relative error from Eq.~(\ref{x0t0r}) is $O(|t|^{-1/9})$. These predicted decay rates are plotted as dashed lines in panels (b) and (c) respectively as well. We can see from these panels that the true decay rates indeed agree with the predictions at large time, which quantitatively confirms Theorem~\ref{Theorem1}.

\begin{figure}[htb]
\begin{center}
\vspace{8.0cm}
\hspace{-2.0cm}
\includegraphics[scale=0.25, bb=300 0 400 360]{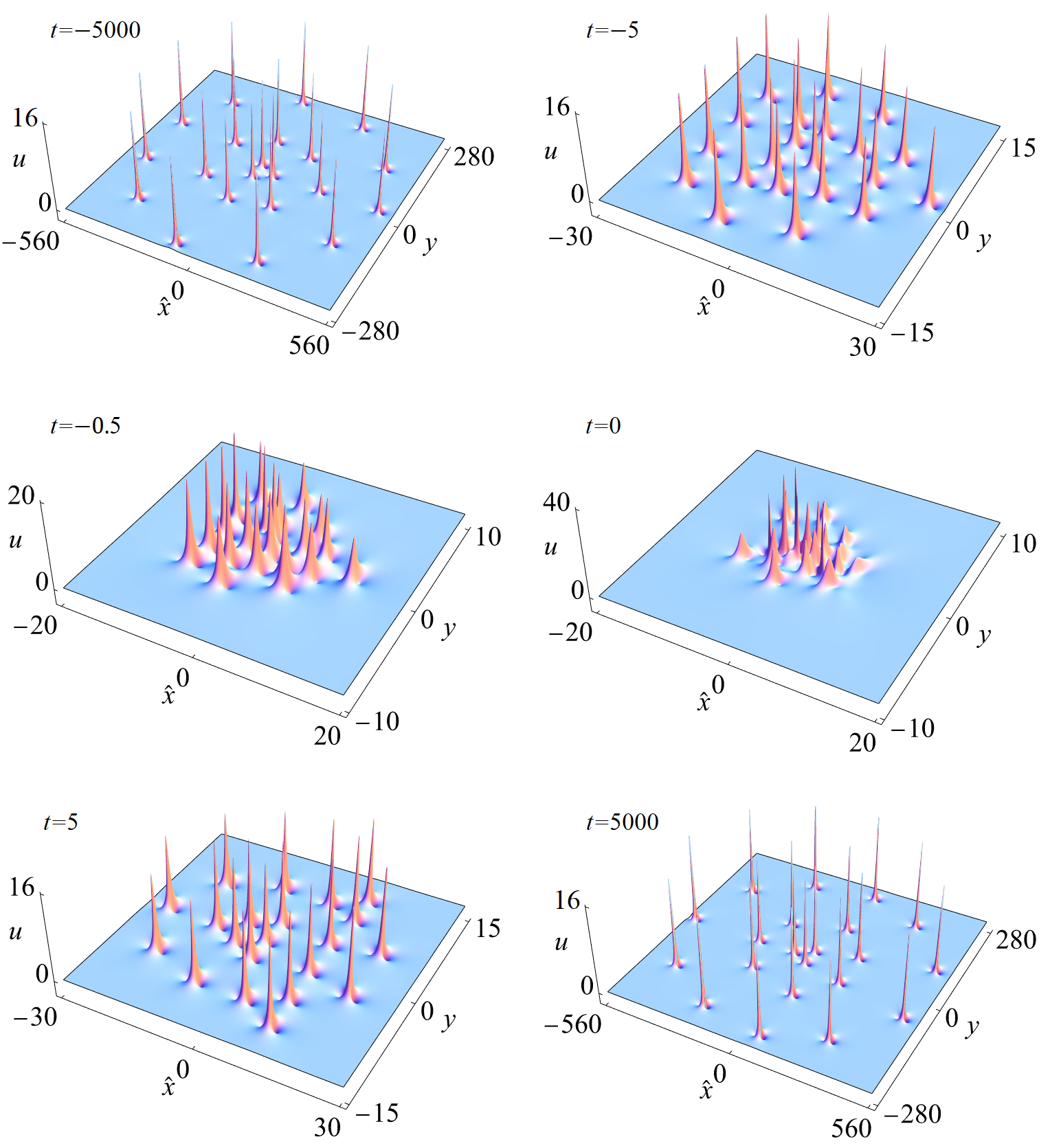}\hspace{2.0cm}
\caption{The true solution $u_{\Lambda} (x,y,t)$ with $\Lambda =(1, 3, 5, 7, 9, 11)$ at time values of $t=-5000, -5, -0.5, 0, 5$ and 5000. Internal parameters are taken as in Eq.~(\ref{para2}), with the other elements of parameter vectors as zero. The axis $\hat{x}=x-12t$ is the moving $x$ coordinate. }  \label{f:fig4}
\end{center}
\end{figure}

\textbf{Example 2.} In our second example, we take $N=6$; so $\Lambda =(1, 3, 5, 7, 9, 11)$. Internal parameters $\textbf{\emph{a}}_1, \dots, \textbf{\emph{a}}_6$ are taken as
\[ \label{para2}
\left.\begin{array}{l}
(a_{1,1}, a_{2,1}, a_{3,1}, a_{4,1}, a_{5,1}, a_{6,1})=(0, 1, 1, 1, -1, -1), \\
(a_{2,2}, a_{3,2}, a_{4,2}, a_{5,2}, a_{6,2})=(1, {\rm{i}}, 2, 3{\rm{i}}, 4),
\end{array}\right\}
\]
with the other elements of parameter vectors $\textbf{\emph{a}}_i \ (1\le i\le 6)$ taken as zero. The true solution from Lemma~1 at six time values of $t=-5000, -5, -0.5, 0,5$ and 5000 is plotted in Fig.~\ref{f:fig4}. It is seen that at $t=-5000$, the solution splits into three concentric rings with 11, 7 and 3 fundamental lumps evenly distributed on them respectively. As time increases to $-5$ and $-0.5$, these 21 fundamental lumps get close to each other, and the shape formed by them has changed as well. As time increases further to 0, these 21 fundamental lumps merge with each other and form several high spikes. As time further increases to $5$, the merged solution splits up into 21 fundamental lumps again. When time continues to increase to 5000, these 21 fundamental lumps evolve into three concentric rings with 11, 7 and 3 fundamental lumps on them again. This lump pattern at $t=5000$ is similar to that at $t=-5000$. However, we should notice that the relative positions of fundamental lumps on the three rings at $t=-5000$ and $5000$ are different (unlike Example 1 in Fig.~\ref{f:fig1}).

\begin{figure}[htb]
\begin{center}
\vspace{-0.0cm}
\includegraphics[scale=0.25, bb=0 0 1150 490]{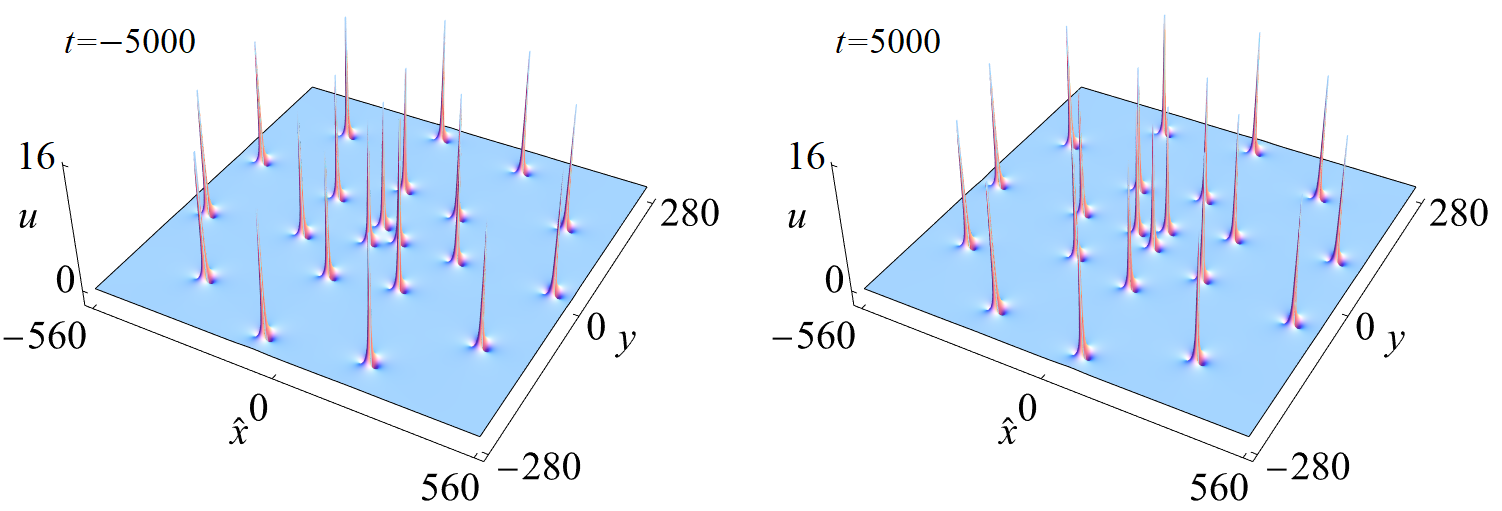}
\caption{Leading-order predictions of lump patterns from Theorem~\ref{Theorem1} for the solution of Fig.~\ref{f:fig4} at large times of $t=\pm 5000$. }  \label{f:fig5}
\end{center}
\end{figure}

Now, we use Theorem~\ref{Theorem1} to analytically predict the solution of Fig.~\ref{f:fig4} at large times of $t=\pm 5000$. Since $N=6$ here, that theorem predicts that the solution of Fig.~\ref{f:fig4} at large times would split into three concentric rings, with 11, 7 and 3 fundamental lumps on them respectively. To determine analytical predictions of lump positions on these rings, we notice from the parameter choices (\ref{para2}) that $\mathbf{D}= \mbox{diag}(0, 1, 1, 1, -1, -1)$. Then, the $3\times 3$ submatrix of $\mathbf{G}$ from Eq.~(\ref{defG}) and its factored $\mathbf{B}$ matrix from Eq.~(\ref{FacAB}) are
\[
\mathbf{G}_{4\le i\le 6, \ 1\le j\le 3}=\left(
\begin{array}{ccc}
 \frac{1}{6} & 0 & 0 \\
 -\frac{1}{8} &  -\frac{1}{3} & -1 \\
 -\frac{3}{40} &  \frac{1}{4} &  \frac{2}{3}
\end{array}
\right),
\quad \mathbf{B}= \left(
\begin{array}{ccc}
 0 & 0 & -\frac{20}{27} \\
 0 & \frac{1}{12} & \frac{1}{9} \\
 \frac{3}{40} & \frac{1}{4} & \frac{2}{3} \end{array}
\right).
\]
This shows that $\beta_{1,1}=\frac{3}{40}$, $\beta_{2,2}=\frac{1}{12}$ and $\beta_{3,3}= -\frac{20}{27}$.  Notice that the above submatrix of $\mathbf{G}$ satisfies our Assumption 1. Using these values, we can obtain leading-order predictions of lump positions on these three rings from Eq.~(\ref{x0t0r}) with $N=6$ and $r=1, 2, 3$. These predicted solutions at large times of $t=\pm 5000$ are plotted in Fig.~\ref{f:fig5}. Comparing these predictions with true solutions at $t=\pm 5000$ in Fig.~\ref{f:fig4}, we can see that the predictions agree with true solutions very well.

\section{Proof of Theorem~\ref{Theorem1} for lump patterns with $\Lambda=(1, 3, \dots, 2N-1)$} \label{secProof}
Now, we prove Theorem~\ref{Theorem1}.

Firstly, we rewrite the determinant $\sigma$ in Eq.~(\ref{Blockmatrix}) as a larger $3N \times 3N$ determinant
\[ \label{3Nby3Ndet2}
\sigma=\left|\begin{array}{cc}
\textbf{O}_{N\times N} & \Phi_{N\times 2N} \\
-\Psi_{2N\times N} & \textbf{I}_{2N \times 2N} \end{array}\right|,
\]
where
\[
\Phi_{i,j}=2^{-(j-1)} S_{2i-j}\left(\textbf{\emph{x}}^{+} + (j-1) \textbf{\emph{s}} +\textbf{\emph{a}}_i\right), \quad \Psi_{i,j}=2^{-(i-1)} S_{2j-i}\left((\textbf{\emph{x}}^{+})^* + (i-1) \textbf{\emph{s}}+\textbf{\emph{a}}_j^*\right).
\]
Performing the Laplace expansion to this larger determinant, we get
\[ \label{sigmanLap}
\sigma=\sum_{1\leq\nu_{1} < \nu_{2} < \cdots < \nu_{N}\leq 2N}\left|
\det_{1 \leq i, j\leq N} \Phi_{i, \nu_j}\right|^2.
\]
When $|t|\gg 1$, $|S_{k}\left(\textbf{\emph{x}}^{+} + \nu \textbf{\emph{s}} +\textbf{\emph{a}}_i\right)|\gg 1$. In this case, the highest $t$-power term of $\sigma$ comes from the index choice of $\nu_{j}=j$ in this Laplace expansion, i.e.,
\[ \label{sigmanLap5}
\sigma\sim \left|\det_{1 \leq i, j\leq N} \Phi_{i, j}\right|^2, \qquad |t|\gg 1.
\]

Now, we analyze the large-time asymptotics of this $\sigma$ determinant for the index vector $\Lambda=(1, 3, \dots, 2N-1)$. For this purpose, we introduce a moving $x$ coordinate
\[ \label{defxhat}
\hat{x}\equiv x-12 t.
\]
Then, the elements $x_{k}^+$ in Eq.~(\ref{defxrp}) become
\[ \label{defx1x3}
x_k^+=\frac{1}{k!}\hat{x}+\frac{2^k}{k!} \textrm{i}y+T_k,
\]
where
\[
T_k\equiv \frac{12(1-3^{k-1})}{k!}t.
\]
In particular,
\[ \label{T2T3}
x_1^+=\hat{x}+2\textrm{i}  y, \quad  T_2=-12t, \quad T_3=-16t.
\]

For $\Lambda=(1, 3, \dots, 2N-1)$, the vector $\textbf{\emph{a}}_i$ has length $n_i=2i-1$. Let us define parameters $c_{i,j}$ by
\[
\sum_{j=0}^{\infty} c_{i,j} \epsilon^j=\exp\left( \sum_{j=1}^{\infty} a_{i,j} \epsilon^j \right),
\]
with $a_{i,j}\equiv 0$ when $j>n_i$. Then, using the definition of Schur polynomials, we get the relation
\[ \label{polyrelation3}
S_k(\textbf{\emph{x}}^{+} + \nu \textbf{\emph{s}}+\textbf{\emph{a}}_i ) = \sum_{j=0}^{k} c_{i,j}S_{k-j}(\textbf{\emph{x}}^{+} + \nu \textbf{\emph{s}}).
\]

When $\sqrt{\hat{x}^2+y^2}=O(|t|^{q}),\ \frac{1}{3} \leq q <\frac{1}{2}$, we have
\[ \label{Skh1}
S_k(\textbf{\emph{x}}^{+} + \nu \textbf{\emph{s}}) \sim  S_k(\textbf{h}),\quad |t|\gg 1,
\]
where $\textbf{h}=(x_1^+, T_2, T_3, 0, 0, \cdots)$.
Splitting $\textbf{h}$ as $(0, T_2, 0, 0, \cdots)+\textbf{v}$, with
\[ \label{defv}
\textbf{v}\equiv (x_1^+, 0, T_3, 0, 0,  \cdots),
\]
we can use the definition of Schur polynomials to show that
\[ \label{polyrelation2}
S_k(\textbf{h}) = \sum_{j=0}^{\left[k/2\right]} \frac{T_2^{j}}{j!} S_{k-2j}(\textbf{v}).
\]

Now, we use the above results to derive the highest-power term of $|t|$ in the $\sigma$ determinant (\ref{3Nby3Ndet2}).

When $1\ll |x_1^+|\ll O(|t|^{1/2})$, we can use relations (\ref{polyrelation3})-(\ref{polyrelation2}) to write the dominant terms of $\Phi$ in Eq.~(\ref{3Nby3Ndet2}) as
\[\label{PhiAsym}
\Phi\sim {\cal F} \mathbf{P}_1 + \mathbf{D} {\cal F} \mathbf{P}_2, \quad |t|\gg 1,
\]
where
\begin{eqnarray}
&& {\cal F} =\left(
             \begin{array}{cccc}
               1 & 0 & \cdots & 0 \\
               T_2 & 1 & \cdots & 0 \\
                \vdots &  \vdots &\ddots & \vdots \\
                \frac{T_2^{N-1}}{\left(N-1\right)!} & \frac{T_2^{N-2}}{\left(N-2\right)!} & \cdots & 1\\
             \end{array}
           \right)_{N\times N},  \\
           &&
           \mathbf{D}= \mbox{diag}\left(c_{1,1}, c_{2,1}, c_{3,1},\cdots, c_{N,1} \right),   \\
           && \mathbf{P}_k= \mbox{Mat}_{1\leq i\leq N,\ 1\leq j\leq 2N} \left( 2^{-(j-1)} S_{2i-j+1-k}(\textbf{v})\right), \quad k=1, 2.   \label{defPk}
\end{eqnarray}
Here, we have dropped certain subdominant terms in $\Phi$ using the fact that for any positive integer $n$ and nonnegative integer $k$,
\[ \label{factS1}
|T_2^{-n}S_{2n+k}(\textbf{v})|\ll |S_{k}(\textbf{v})|, \quad 1\ll |x_1^+|\ll O(|t|^{1/2}).
\]
Note that this fact would not hold when $|x_1^+|=O(1)$ or $|x_1^+|\ge O(|t|^{1/2})$, with $T_2^{-1}S_3(\textbf{v})=O(S_1(\textbf{v}))$ when $|x_1^+|=O(1)$ as an example. The above $\mathbf{D}$ matrix can be seen the same as that defined in Eq.~(\ref{defD}) since $c_{i,1}=a_{i,1}$. The above ${\cal F}$ matrix can be rewritten as
\[
{\cal F}=\mathbf{E}^{-1} \mathbf{F} \mathbf{E},
\]
where $\mathbf{E}\equiv \mbox{diag}\left(1, T_2^{-1}, \cdots, T_2^{-(N-1)} \right)$, and $\mathbf{F}$ is as defined in Eq.~(\ref{defF}).
Then we have ${\cal F}^{-1} \mathbf{D} {\cal F} = \mathbf{E}^{-1} \mathbf{G} \mathbf{E}$,
where $\mathbf{G}$ is as defined in Eq.~(\ref{defG}). Thus, $\Phi$'s asymptotics (\ref{PhiAsym}) becomes
\[\label{PhiAsymF}
\Phi\sim {\cal F} \left[ \mathbf{P}_1 + \mathbf{E}^{-1} \mathbf{G} \mathbf{E} \mathbf{P}_2 \right], \quad |t|\gg 1.
\]
This $\Phi$ can be written out explicitly as
\small
\begin{eqnarray} \label{PhiMatrixForm}
\Phi\sim  {\cal F} \left(\begin{array}{lll}
S_1(\textbf{v}) + g_{1,1}S_0(\textbf{v})   & 2^{-1} S_0(\textbf{v}) & \cdots \\
S_3(\textbf{v}) +  T_2 g_{2,1} S_0(\textbf{v}) + g_{2,2} S_2(\textbf{v})  & 2^{-1} \left( S_2(\textbf{v}) + g_{2,2} S_1(\textbf{v}) \right) & \cdots \\
\vdots & \vdots & \vdots \\
S_{2N-3}(\textbf{v}) + T_2^{N-2} g_{N-1,1} S_0(\textbf{v}) + T_2^{N-3}  g_{N-1,2}S_2(\textbf{v}) +\cdots &
2^{-1} \left(S_{2N-4}(\textbf{v})+T_2^{N-3} g_{N-1,2} S_1(\textbf{v})  +\cdots\right) &
\cdots \\
S_{2N-1}(\textbf{v}) + T_2^{N-1} g_{N,1}S_0(\textbf{v}) + T_2^{N-2} g_{N,2}S_2(\textbf{v})+\cdots &
2^{-1} \left( S_{2N-2}(\textbf{v}) + T_2^{N-2} g_{N,2}S_1(\textbf{v}) +\cdots \right) &
\cdots
\end{array}\right).
\end{eqnarray}
\normalsize

\subsection{The outer-most ring case ($r=1$)}
We first prove Theorem~\ref{Theorem1} for the outer-most ring of fundamental lumps where $r=1$.

In this case, under Assumption 1, $g_{N, 1}\neq 0$ since $g_{N,1}=M_1$. In the last row and first column of the right matrix of the above equation (\ref{PhiMatrixForm}), we balance its first two terms $S_{2N-1}(\textbf{v})$ and  $T_2^{N-1} g_{N,1}S_0(\textbf{v})$ as the leading-order terms. When $x_1^{+}=O(|t|^{q})$ with $q\ge 1/3$, $S_{k}(\textbf{v})=O\left((x_1^{+})^k\right)$. Thus, this balance gives $x_1^+=O\left(|t|^{\frac{N-1}{2N-1}}\right)$. Notice that $\frac{1}{3}\le \frac{N-1}{2N-1}<\frac{1}{2}$ for $N>1$. In this $(x, y)$ region, it is easy to check that all the other $g_{ij}$ terms in (\ref{PhiMatrixForm}) are subdominant, and we have
\[ \label{PhiMatrixForm2}
\Phi\sim {\cal F}\left(\begin{array}{lll}
S_1(\textbf{v})  & 2^{-1}S_0(\textbf{v}) & \cdots \\
S_3(\textbf{v})   & 2^{-1}S_2(\textbf{v}) & \cdots \\
\vdots & \vdots & \vdots \\
S_{2N-3}(\textbf{v}) & 2^{-1} S_{2N-4}(\textbf{v}) & \cdots \\
S_{2N-1}(\textbf{v}) + T_2^{N-1} g_{N,1} S_{0}(\textbf{v}) &
2^{-1} S_{2N-2}(\textbf{v})  & \cdots
\end{array}\right)_{N\times 2N}.
\]
For any $x_1^+=O\left(|t|^{q}\right)$ with $q>1/3$,
\[ \label{Sk1}
S_k(\textbf{v})\sim \frac{1}{k!}(x_1^+)^k.
\]
Thus, if $\frac{N-1}{2N-1}>\frac{1}{3}$, i.e., $N>2$, then using this $S_k(\textbf{v})$ asymptotics and introducing the scaled variable
\[\label{defz11}
z\equiv T_2^{-\frac{N-1}{2N-1}}x_1^+ =T_2^{-\frac{N-1}{2N-1}}\left(\hat{x}+2\textrm{i}  y\right),
\]
we can see from Eq~(\ref{PhiMatrixForm2}) that
\[ \label{detPhiasym1}
\det_{1 \leq i, j\leq N} \Phi_{i, j} \sim \gamma_1 T_2^{\rho_1} Q_{N}(z; \beta_{1, 1}),
\]
where the polynomial $Q_n(z; \beta)$ for an arbitrary integer $n$ and complex number $\beta$ is defined as
\[ \label{defH1}
Q_n(z; \beta)\equiv \det \left(\begin{array}{llll}
 z & 1 & \cdots  & 0          \\
\frac{z^3}{3!} & \frac{z^2}{2!} & \cdots  & 0 \\
\vdots & \vdots &   \vdots & \vdots \\
\frac{z^{2n-3}}{(2n-3)!}  & \frac{z^{2n-4}}{(2n-4)!} & \cdots  & \frac{z^{n-2}}{(n-2)!} \\
\frac{z^{2n-1}}{(2n-1)!} + \beta  & \frac{z^{2n-2}}{(2n-2)!} & \cdots  & \frac{z^{n}}{n!}  \end{array} \right)_{n\times n},
\]
$\gamma_1=2^{-N(N-1)/2}$, and $\rho_1=(N-1)N(N+1)/2(2N-1)$. This $Q_n(z; \beta)$ polynomial can be calculated as \cite{OhtaJY2012}
\[ \label{Qnz}
Q_n(z; \beta)=\frac{z^{n(n+1)/2}}{\kappa_n}+\beta \frac{z^{(n-2)(n-1)/2}}{\kappa_{n-2}},
\]
where $\kappa_{n}= \prod_{j=1}^{n}(2j-1)!!$, and its nonzero roots are given by the equation
\[
z^{2n-1}=-\beta (2n-1)!!(2n-3)!!,
\]
i.e., they are the $(2n-1)$-th roots of $-\beta (2n-1)!!(2n-3)!!$.

Substituting the asymptotics (\ref{detPhiasym1}) into the $\sigma$ asymptotics (\ref{sigmanLap5}), we get
\[ \label{sigmaasym1}
\sigma \sim \gamma_1^2 \hspace{0.05cm} |T_2|^{2\rho_1} \left|Q_N(z; \beta_{1,1}) \right|^2.
\]
Combining this $\sigma$ asymptotics with Eq.~(\ref{Schpolysolu}), we see that when $x_1^+=O(|t|^{\frac{N-1}{2N-1}})$, i.e., when $\sqrt{\hat{x}^2+y^2}=O(|t|^{\frac{N-1}{2N-1}})$, the solution $u_\Lambda(x,y,t)$ would be asymptotically zero in this region at large $|t|$, except when $(\hat{x}, y)$ is at or near the location $\left(\tilde{x}_0, \tilde{y}_0\right)$, where
\[
z_0=T_2^{-\frac{N-1}{2N-1}}\left(\tilde{x}_0 +2\textrm{i} \tilde{y}_0\right)
\]
is a nonzero root of $Q_N(z; \beta_{1,1})$, i.e., $z_0$ is one of the $(2N-1)$-th roots of $-\beta_{1,1}(2N-1)!!(2N-3)!!$.
This $z_0$ value matches the one given in Theorem~\ref{Theorem1} for $r=1$.

Next, we show that when $(\hat{x}, y)$ is at or near these $\left(\tilde{x}_0, \tilde{y}_0\right)$ locations, the solution $u_{\Lambda} (x,y,t)$ in Lemma 1 asymptotically reduces to a fundamental lump. Near these locations, the asymptotics (\ref{sigmaasym1}) would break down since its $Q_N(z; \beta_{1,1})$ term is near zero. Thus, to obtain the correct $\sigma$ asymptotics there, we need to calculate the next-order terms. By reviewing our prior asymptotics, we can see that in these $(\hat{x}, y)$ regions where $x_1^+=O(|t|^{\frac{N-1}{2N-1}})$, the terms which we have neglected in $\Phi$'s asymptotics (\ref{PhiMatrixForm2}) are terms of relative order $O(|t|^{-\frac{1}{2N-1}})$. Including such terms, we can see that the full asymptotics of $\sigma$'s Laplace expansion (\ref{sigmanLap}) is
\[ \label{sigmaasym2}
\sigma =\gamma_1 \hspace{0.05cm} |T_2|^{\rho_1} \left\{
\left|Q_N(z; \beta_{1,1})+ O\left(T_2^{-\frac{1}{2N-1}}\right)\right|^2
+\frac{1}{4} \left|Q_N'(z; \beta_{1,1})+ O\left(T_2^{-\frac{1}{2N-1}}\right)\right|^2 T_2^{-\frac{2(N-1)}{2N-1}}+O\left(T_2^{-\frac{4(N-1)}{2N-1}}\right)
\right\},
\]
where the middle term in the above asymptotics comes from the Laplace expansion (\ref{sigmanLap}) for the index $(\nu_1, \nu_2, \dots, \nu_N)$ being $(1, 2, \dots, N-1, N+1)$, and the prime denotes differentiation with respect to $z$. Expanding the $Q_N(z; \beta_{1,1})$ term around its nonzero root $z_0$, and replacing $Q_N'(z; \beta_{1,1})$ by its dominant term $Q_N'(z_0; \beta_{1,1})$ which is nonzero since each of the nonzero roots $z_0$ in $Q_N(z; \beta_{1,1})$ is simple, we get
\begin{eqnarray}\label{sigmaasym3}
&& \sigma = \gamma_1 \hspace{0.05cm} |T_2|^{\rho_1-\frac{2(N-1)}{2N-1}} \left|Q_N'(z_0; \beta_{1,1})\right|^2 \left\{\left|  (\hat{x}-\tilde{x}_0)+2{\rm{i}}(y-\tilde{y}_0)+O\left(T_2^{\frac{N-2}{2N-1}}\right) \right|^2 +\frac{1}{4} + O\left(T_2^{-\frac{1}{2N-1}}\right)\right\}.
\end{eqnarray}
The above equation can be rewritten as
\[
\sigma =  \gamma_1 \hspace{0.05cm} |T_2|^{\rho_1-\frac{2(N-1)}{2N-1}} \left|Q_N'(z_0; \beta_{1,1})\right|^2 \left(\left|  (\hat{x}-x_0)+2{\rm{i}}(y-y_0)\right|^2+\frac{1}{4}\right)\left(1+O\left(T_2^{-\frac{1}{2N-1}}\right)\right),
\]
where
\[ \label{x0y05}
x_0+2{\rm{i}}y_0=\tilde{x}_0+2{\rm{i}}\tilde{y}_0+O\left(T_2^{\frac{N-2}{2N-1}}\right)=z_0 T_2^{\frac{N-1}{2N-1}}\left(1+O\left(|T_2|^{-\frac{1}{2N-1}}\right)\right).
\]
Inserting this $\sigma$ asymptotics into Eq.~(\ref{Schpolysolu}), we see that the resulting solution $u_{\Lambda} (x,y,t)$ is
\[
u_{\Lambda} (x,y,t)=u_1(x-x_0, y-y_0)+O\left(|t|^{-\frac{1}{2N-1}}\right),
\]
which is asymptotically a fundamental lump located at the $(x_0, y_0)$ positions given in Eq.~(\ref{x0y05}). This proves Theorem~\ref{Theorem1} for the case of $r=1$.

\subsection{The second outer-most ring case ($r=2$)}
The outer-most ring case with $r=1$ above pertains to $x_1^+=O\left(|t|^{\frac{N-1}{2N-1}}\right)$. Now, we consider the case of $r=2$ in Theorem~\ref{Theorem1}, where $x_1^+=O\left(|t|^{q}\right)$ with $q=\frac{N-3}{2N-5}>\frac{1}{3}$. Notice that this $q$ is less than $1/2$. This case corresponds to the second outer-most ring of fundamental lumps. In this $(x, y)$ region, the matrix elements in the last two rows and the first column of the right matrix of Eq.~(\ref{PhiMatrixForm}) have the following asymptotics
\[\label{secondeadring2}
S_{2N-3}(\textbf{v}) + T_2^{N-2} g_{N-1,1} S_{0}(\textbf{v}) + T_2^{N-3} g_{N-1,2} S_{2}(\textbf{v})+\cdots = T_2^{N-2} g_{N-1,1} S_{0}(\textbf{v})\left(1+\mathcal{O}(|t|^{-\frac{1}{2N-5}})\right),
\]
and
\[\label{secondeadring1}
S_{2N-1}(\textbf{v}) + T_2^{N-1} g_{N, 1} S_{0}(\textbf{v}) + T_2^{N-2} g_{N,2} S_{2}(\textbf{v})+\cdots = T_2^{N-1} g_{N,1} S_{0}(\textbf{v})\left(1+\mathcal{O}(|t|^{-\frac{1}{2N-5}})\right),
\]
while the upper $N-2$ rows and first column of that matrix are dominated by their first term $S_{2j-1}(\textbf{v})$. Using (\ref{secondeadring1}) and row operations to eliminate the leading-order term $T_2^{N-2} g_{N-1,1} S_{0}(\textbf{v})$ of (\ref{secondeadring2}) in Eq.~(\ref{PhiMatrixForm}) and dropping subdominant terms, we find that
\begin{eqnarray}
&& \det_{1 \leq i, j\leq N} \Phi_{i, j} \sim \gamma_1
\det \left(\begin{array}{llll}
S_1(\textbf{v})  & S_0(\textbf{v}) & 0 &\cdots \\
S_3(\textbf{v}) & S_2(\textbf{v})  & S_1(\textbf{v})  &\cdots \\
\vdots & \vdots & \vdots &\vdots \\
S_{2N-5}(\textbf{v})   &
S_{2N-6}(\textbf{v}) &S_{2N-7}(\textbf{v}) &
\cdots \\
S_{2N-3}(\textbf{v}) + T_2^{N-3} \beta_{2,2} S_2(\textbf{v}) &
S_{2N-4}(\textbf{v})  + T_2^{N-3} \beta_{2,2} S_{1}(\textbf{v})& S_{2N-5}(\textbf{v}) + T_2^{N-3}\beta_{2,2}S_{0}(\textbf{v}) &\cdots \\
T_2^{N-1}\beta_{1,1}S_{0}(\textbf{v}) & 0 & 0 & \cdots
\end{array}\right)_{N\times N}  \nonumber  \\
&& \hspace{1cm} = \gamma_1\beta_{1,1}T_2^{N-1} \det \left(\begin{array}{llll}
S_1(\textbf{v})  & S_0(\textbf{v}) & 0 &\cdots \\
S_3(\textbf{v})  & S_2(\textbf{v}) &  S_1(\textbf{v}) &\cdots \\
\vdots & \vdots & \vdots &\vdots \\
S_{2N-7}(\textbf{v}) &S_{2N-8}(\textbf{v}) & S_{2N-9}(\textbf{v}) &
\cdots \\
S_{2N-5}(\textbf{v}) + T_2^{N-3} \beta_{2,2} S_0(\textbf{v}) &
S_{2N-6}(\textbf{v})  & S_{2N-7}(\textbf{v}) & \cdots
\end{array}\right)_{(N-2)\times (N-2)},    \label{Phiasym5}
\end{eqnarray}
where $\beta_{2,2}= g_{N-1,2}- (g_{N-1,1}/g_{N,1})g_{N,2}=M_2/M_1$, and $M_r$ is as defined in Eq.~(\ref{defMr0}). This $\beta_{2,2}$ is the same $\beta_{2,2}$ as in the $\mathbf{A}\mathbf{B}$ factorization (\ref{FacAB}) and is nonzero under Assumption~1. Since $x_1^+=O\left(|t|^{q}\right)$ with $q>1/3$, then using the $S_k(\textbf{v})$ asymptotics (\ref{Sk1}), we can employ similar techniques as used earlier to reduce the above asymptotics to
\[
\det_{1 \leq i, j\leq N} \Phi_{i, j} \sim \gamma_2 T_2^{\rho_2} Q_{N-2}(z; \beta_{2, 2}),
\]
where $z$ is the scaled variable
\[\label{defz12}
z\equiv T_2^{-\frac{N-3}{2N-5}}x_1^+ =T_2^{-\frac{N-3}{2N-5}}\left(\hat{x}+2\textrm{i}  y\right),
\]
$\gamma_2=\gamma_1\beta_{1,1}$, and $\rho_2=N-1+(N-3)(N-2)(N-1)/2(2N-5)$. Then, the $\sigma$ asymptotics (\ref{sigmanLap5}) becomes
\[ \label{sigmaasym6}
\sigma \sim \gamma_2^2 \hspace{0.05cm} |T_2|^{2\rho_2} \left|Q_{N-2}(z; \beta_{2,2}) \right|^2.
\]
Because of this, when $\sqrt{\hat{x}^2+y^2}=O(|t|^{\frac{N-3}{2N-5}})$, the solution $u_\Lambda(x,y,t)$ would be asymptotically zero in this region at large $|t|$, except when $(\hat{x}, y)$ is at or near the location $\left(\tilde{x}_0, \tilde{y}_0\right)$, where
\[
z_0=T_2^{-\frac{N-3}{2N-5}}\left(\tilde{x}_0 +2\textrm{i} \tilde{y}_0\right)
\]
is a nonzero root of $Q_{N-2}(z; \beta_{2, 2})$, i.e., $z_0$ is one of the $(2N-5)$-th roots of $-\beta_{2,2}(2N-5)!!(2N-7)!!$. This $z_0$ value matches the one given in Theorem~\ref{Theorem1} for $r=2$.

Repeating calculations similar to that for the $r=1$ case, we can further show that when $\sqrt{\hat{x}^2+y^2}=O(|t|^{\frac{N-3}{2N-5}})$,
\[
u_{\Lambda} (x,y,t)=u_1(x-x_0, y-y_0)+O\left(|t|^{-\frac{1}{2N-5}}\right),
\]
where
\[ \label{x0y09}
x_0+2{\rm{i}}y_0=z_0 T_2^{\frac{N-3}{2N-5}}\left(1+O\left(|t|^{-\frac{1}{2N-5}}\right)\right).
\]
That is, the solution $u_{\Lambda} (x,y,t)$ asymptotically reduces to a fundamental lump located at the above $(x_0, y_0)$ positions. This proves Theorem~\ref{Theorem1} for $r=2$.

\subsection{The case of the inner-most ring with three fundamental lumps ($r=[N/2]$ for even $N$)}
For higher $r$ values corresponding to rings closer to the ring center, the proof of Theorem~\ref{Theorem1} proceeds with little modification. We use lower rows of $g_{i,j}$ terms in the right matrix of (\ref{PhiMatrixForm}) to eliminate certain $g_{i,j}$ terms of small $j$ in the higher rows of that matrix, which is equivalent to the $\mathbf{A}\mathbf{B}$ factorization (\ref{FacAB}). Then, when $x_1^+=O(|t|^q)$ with $q=\frac{N-1-2(r-1)}{2N-1-4(r-1)}>1/3$, we can reduce (\ref{PhiMatrixForm}) to a simpler matrix so that $\det_{1 \leq i, j\leq N} \Phi_{i, j}$ is asymptotically proportional to $Q_{N-2r+2}(z; \beta_{r, r})$, with $z$ being a certain scaled $x_1^+$ variable. This would yield the results in Theorem~\ref{Theorem1} for higher $r$.

This process proceeds until we reach the $r=[N/2]$-th ring for even $N$, in which case $\frac{N-1-2(r-1)}{2N-1-4(r-1)}=\frac{1}{3}$. In this case, $x_1^+=O(|t|^{1/3})$. This case is special because the $S_k(\textbf{v})$ asymptotics (\ref{Sk1}) does not hold here. Indeed, now
\[ \label{Sk2}
S_3(\textbf{v})=\frac{(x_1^+)^3}{3!}+T_3,
\]
where the two terms in it are of the same order in $|t|$. In this special case, much of the previous calculations still holds, until we reach the following reduced asymptotics
\begin{eqnarray} \label{detPhiasym8}
\det_{1 \leq i, j\leq N} \Phi_{i, j} \sim \gamma_{[N/2]} T_2^{\hat{\rho}} \det \left(\begin{array}{ll}
S_1(\textbf{v})  & S_0(\textbf{v}) \\
S_3(\textbf{v})+T_2 \beta_{[N/2], [N/2]} S_0(\textbf{v})  & S_2(\textbf{v})
\end{array}\right),
\end{eqnarray}
which is the counterpart of Eq.~(\ref{Phiasym5}). Here, $\gamma_{[N/2]}=\gamma_1\prod_{r=1}^{[N/2]-1}\beta_{r,r}$, and $\hat{\rho}=(N^2-4)/4$. Since
\[
S_0(\textbf{v})=1, \quad S_1(\textbf{v})=x_1^+, \quad S_2(\textbf{v})=\frac{(x_1^+)^2}{2!},
\]
introducing the scaled variable
 \[\label{defz13}
z\equiv T_2^{-\frac{1}{3}}x_1^+ =T_2^{-\frac{1}{3}}\left(\hat{x}+2\textrm{i}  y\right),
\]
we can easily find that the asymptotics (\ref{detPhiasym8})
becomes
\begin{eqnarray} \label{detPhiasym9}
\det_{1 \leq i, j\leq N} \Phi_{i, j} \sim \gamma_{[N/2]} T_2^{\hat{\rho}+1} \det \left(\begin{array}{ll}
z  & 1 \\
\frac{z^3}{3!}+\beta_{[N/2], [N/2]}+\frac{4}{3}  & \frac{z^2}{2!}
\end{array}\right)=\gamma_{[N/2]} T_2^{\hat{\rho}+1}Q_{2}\left(z; \beta_{[N/2], [N/2]}+4/3\right).
\end{eqnarray}
Here the factor $4/3$ comes from the ratio of $T_3/T_2$. Then, the $\sigma$ asymptotics (\ref{sigmanLap5}) becomes
\[
\sigma \sim \gamma_{[N/2]}^2 \hspace{0.05cm} |T_2|^{2(\hat{\rho}+1)} \left|Q_{2}\left(z; \beta_{[N/2], [N/2]}+4/3\right) \right|^2.
\]
This asymptotics shows that, when $\sqrt{\hat{x}^2+y^2}=O(|t|^{\frac{1}{3}})$, the solution $u_\Lambda(x,y,t)$ would be asymptotically zero in this region at large $|t|$, except when $(\hat{x}, y)$ is at or near the location $\left(\tilde{x}_0, \tilde{y}_0\right)$, where
\[
z_0=T_2^{-\frac{1}{3}}\left(\tilde{x}_0 +2\textrm{i} \tilde{y}_0\right)
\]
is a nonzero root of $Q_{2}\left(z; \beta_{[N/2], [N/2]}+4/3\right)$, i.e., $z_0$ is one of the cubic roots of $-(\beta_{[N/2],[N/2]}+4/3)3!!1!!$. There are three such $\left(\tilde{x}_0, \tilde{y}_0\right)$ locations since there are three such $z_0$ roots. Similarly as before, we can further show that at each of these three locations, the solution $u_{\Lambda} (x,y,t)$ asymptotically reduces to a fundamental lump located at the $(x_0, y_0)$ positions specified in Theorem~\ref{Theorem1} for the $r=[N/2]$-th ring with even $N$. This finishes the proof of Theorem~\ref{Theorem1}.

\section{Lump patterns at large times for $\Lambda\ne (1, 3, \dots, 2N-1)$} \label{sec5}
In this section, we discuss patterns of higher order lumps at large times for $\Lambda\ne (1, 3, \dots, 2N-1)$ under general internal parameters $\{\textbf{\emph{a}}_{i}\}$ where $a_{i,1}$ is dependent on the $i$ index.

Let us first recall from Ref.~\cite{YangYangKPI} that if $a_{i,j}$ is independent of the $i$ index, when $\Lambda= (1, 3, \dots, 2N-1)$, the solution pattern would comprise fundamental lumps arranged in triangular shapes, which are described by root structures of the Yablonskii-Vorob'ev polynomials; and when $\Lambda\ne (1, 3, \dots, 2N-1)$, the solution pattern would comprise fundamental lumps arranged in nontriangular shapes in the outer region, which are described by nonzero-root structures of the associated Wronskian-Hermit polynomials, together with possible fundamental lumps arranged in triangular shapes in the inner region, which are described by root structures of the Yablonskii-Vorob'ev polynomials.

Now, if $a_{i,1}$ is dependent on the $i$ index, we have shown in previous sections that when $\Lambda= (1, 3, \dots, 2N-1)$, the solution pattern would generically comprise fundamental lumps uniformly distributed on concentric rings. When $\Lambda\ne (1, 3, \dots, 2N-1)$, it turns out that the solution pattern would comprise fundamental lumps arranged in nontriangular shapes in the outer region, which are described by nonzero-root structures of the associated Wronskian-Hermit polynomials, together with possible fundamental lumps generically located on concentric rings in the inner region. The analogy of this result to that in \cite{YangYangKPI} is clear, except that triangular shapes in \cite{YangYangKPI} are replaced by concentric-ring shapes now.

To describe our results when $\Lambda\ne (1, 3, \dots, 2N-1)$, let us first remind the reader about Wronskian-Hermit polynomials. Let $q_{k}(z)$  be polynomials defined by
\[\label{qk}
\sum_{k=0}^{\infty} q_{k}(z) \epsilon^k= \exp \left(z \epsilon + \epsilon^2 \right).
\]
These $q_{k}(z)$ polynomials are related to Hermit polynomials through simple variable scalings. Then, for any positive integer $N$ and index vector $\Lambda=(n_1, n_2, \dots, n_N)$, where $\{n_i\}$ are positive and distinct integers, the Wronskian-Hermite polynomial $W_{\Lambda}(z)$ is defined as the Wronskian of $q_k(z)$ polynomials
\[\label{WronskianHermite}
W_{\Lambda}(z)=\mbox{Wron}\left[ q_{n_1}(z), q_{n_2}(z),\ldots, q_{n_N}(z) \right].
\]
An important property of these Wronskian-Hermite polynomials is that the shape formed by their nonzero roots is non-triangular due to the quartet symmetry of their roots. In addition, the multiplicity of the zero root in $W_{\Lambda}(z)$ is a triangular number $d(d+1)/2$, where
\[ \label{defd}
d\equiv \left\{
\begin{array}{ll} k_{odd}-k_{even}, &  \mbox{when} \hspace{0.1cm} k_{odd}-k_{even} \ge 0, \\  k_{even}-k_{odd}-1, & \mbox{when} \hspace{0.1cm} k_{odd}-k_{even} \le -1, \end{array}
\right.
\]
and $k_{odd}$, $k_{even}$ are the numbers of odd and even elements in the index vector $(n_1, n_2, \dots, n_N)$ respectively.

Regarding nonzero roots of Wronskian-Hermite polynomials, it was conjectured that they are all simple \cite{Felder2012}. If this conjecture holds, then the Wronskian-Hermite polynomial $W_{\Lambda}(z)$ would contain
\[ \label{defNW}
N_{W}=\sum_{i=1}^N n_i -\frac{N(N-1)}{2}-\frac{d(d+1)}{2}
\]
simple roots, where $\sum_{i=1}^N n_i -\frac{N(N-1)}{2}$ is the degree of the Wronskian-Hermite polynomial $W_{\Lambda}(z)$ \cite{YangYangKPI}.

\subsection{Main results}

Our results on solution patterns of higher order lumps at large times for $\Lambda\ne (1, 3, \dots, 2N-1)$ can now be described as follows.

\begin{thm} \label{Theorem2}
If $\Lambda\ne (1, 3, \dots, 2N-1)$ for any positive integer $N$, $d>1$ where $d$ is as defined in Eq.~(\ref{defd}), and nonzero roots of Wronskian-Hermite polynomials are all simple, then when $|t|\gg 1$, the higher-order lump solution $u_{\Lambda} (x,y,t)$ in Lemma 1 would split into $N_W$ fundamental lumps in the outer region, and $d(d+1)/2$ fundamental lumps in the inner region under Assumption 2 of the next subsection.
\begin{enumerate}
\item In the outer region, the $(x_0, y_0)$ positions of these $N_W$ fundamental lumps $u_{1}(x-x_{0}, \hspace{0.04cm}  y-y_{0}, t)$ are given by the equation
\[\label{x0t0r8}
x_0+2{\rm{i}}y_0= z_0 \hspace{0.05cm} (-12t)^{\frac{1}{2}}\left(1+O\left(|t|^{-\frac{1}{2}}\right)\right),
\]
where $z_0$ is each of the $N_W$ nonzero simple roots of $W_{\Lambda}(z)$.
\item In the inner region, under Assumption 2 of the next subsection (which holds generically for general internal parameter values), these $d(d+1)/2$ fundamental lumps will be located on $[d/2]$ concentric rings centered at $(x, y)=(12t, 0)$, with one of them also located in the $O(1)$ neighborhood of the ring center when $d$ is odd. The $r$-th ring (counting from outside with $1\le r\le [d/2]$) contains $2d-1-4(r-1)$ fundamental lumps $u_{1}(x-x_{0}, \hspace{0.04cm}  y-y_{0}, t)$, and their $(x_0, y_0)$ positions (relative to the ring center $(x, y)=(12t, 0)$) are given by the equation
\[\label{x0t0r9}
x_0+2{\rm{i}}y_0= z_0 \hspace{0.05cm} (-12t)^{\frac{d-1-2(r-1)}{2d-1-4(r-1)}}\left(1+O\left(|t|^{-\frac{1}{2d-1-4(r-1)}}\right)\right),
\]
where $z_0$ is every one of the $\left(2d-1-4(r-1)\right)$-th roots of $-\hat{\beta}_{r,r}\left(2d-1-4(r-1)\right)!!\left(2d-3-4(r-1)\right)!!$, except in the case of even $d$ and on its $[d/2]$-th (inner-most) ring, in which case $z_0$ is every one of the cubic roots of $-(\hat{\beta}_{[d/2],[d/2]}+4/3)3!!1!!$. Here, $\{\hat{\beta}_{r,r}, 1\le r\le [r/2]\}$ are complex parameters whose formulae will be provided in the next subsection. Written mathematically, we have the following solution asymptotics
\[
u_{\Lambda} (x,y,t)=u_1(x-x_0, y-y_0)+O\left(|t|^{-\frac{1}{2d-1-4(r-1)}}\right), \quad |t|\gg 1,
\]
where $1\le r\le [d/2]$, and $(x_0, y_0)$ is as given above for each of the stated roots $z_0$.
\end{enumerate}
\end{thm}
The proof of this theorem will be given in the next section.

This theorem indicates that when $\Lambda\ne (1, 3, \dots, 2N-1)$, the solution pattern in the outer region of $O(|t|^{1/2})$ is the same as that in the special case of $a_{i,j}$ being independent of the index $i$ as considered in Ref.~\cite{YangYangKPI}. The reason is obviously that the leading-order prediction formulae (\ref{x0t0r8}) for locations of fundamental lumps in the outer region do not depend on the $a_{i,j}$ parameter values. In other words, this same leading-order outer-region prediction applies to all higher-order lump solutions. As we have seen in the special case of \cite{YangYangKPI}, this outer pattern is described by the nonzero root structure of the underlying Wronskian-Hermite polynomial and is non-triangular. As time moves from large negative to large positive, this non-triangular pattern switches its $x$ and $y$ directions. In addition, fundamental lumps in this outer region separate from each other in proportion to $|t|^{1/2}$ at large time. Since this outer-region's leading-order prediction for general higher-order lumps is the same as that for special higher-order lumps detailed in \cite{YangYangKPI}, we will not give much attention to this part of the prediction in this paper.

The prediction of Theorem~\ref{Theorem2} for the inner region shows that, in this inner region of $O(|t|^q)$ with $\frac{1}{3}\le q< \frac{1}{2}$, the solution pattern is similar to that stated in Theorem~\ref{Theorem1}, i.e., the pattern would generically comprise fundamental lumps uniformly distributed on concentric rings, plus another fundamental lump near the ring center when $d$ is odd. The main difference from Theorem~\ref{Theorem1} is only that the number $N$ would be changed to $d$, and the numbers $\beta_{r,r}$ would be changed to $\hat{\beta}_{r,r}$. In this inner region, fundamental lumps separate from each other in proportion to $|t|^{\frac{m}{2m+1}}$, where $m$ is a certain positive integer that is different on different rings. By choosing the $d$ and $r$ values properly, we can get separation rates of $|t|^{\frac{m}{2m+1}}$ for any positive integer $m$ in this inner region.

Theorem~\ref{Theorem2} assumed that $d>1$. If $d=0$, there would not be any fundamental lumps in the inner region. If $d=1$, there would be a single fundamental lump located in the $O(1)$ neighborhood of the wave center $(x, y)=(12t, 0)$. In both cases, there are no rings of lumps in the inner region.

\subsection{The assumption and $\hat{\beta}_{r,r}$ values in Theorem~\ref{Theorem2}} \label{secAssumptions}
Theorem~\ref{Theorem2} was stated under some assumptions. It also involved some complex parameters $\{\hat{\beta}_{r,r}, 1\le r\le [r/2]\}$. We present these assumptions and provide formulae for $\hat{\beta}_{r,r}$ in this subsection.

When $d>1$, we can see from $d$'s definition (\ref{defd}) that there are two cases:  $k_{odd}-k_{even}>1$ and $k_{even}-k_{odd}>2$. In the former case, there are more odd indices in $\Lambda$; while in the latter case, there are more even indices in $\Lambda$.

Before presenting our assumptions, we need to introduce some notations which would differ for these two cases.

\subsubsection{Notations for the case of $k_{odd}-k_{even}>1$} \label{seccase1}
In this case, there are more odd indices than even indices in $\Lambda$, and $d=k_{odd}-k_{even}$, where $d$ is as defined in Eq.~(\ref{defd}). We group such indices such that $n_1<n_2<\cdots<n_{k_{odd}}$ are all odd indices, followed by $\hat{n}_{1}<\hat{n}_{2}< \cdots< \hat{n}_{k_{even}}$ which are all even indices, with $k_{odd}+k_{even}=N$. That is, $\Lambda=(n_1, n_2, \dots, n_{k_{odd}},\hat{n}_{1}, \hat{n}_{2}, \dots, \hat{n}_{k_{even}})$. This index grouping does not affect the higher-order lump solution in Lemma~1.

We define four matrices
\begin{eqnarray}
&& \mathbf{D}_1= \mbox{diag}\left(a_{1,1}, a_{2,1}, \cdots, a_{k_{odd},1} \right), \label{defD1} \\
&& \mathbf{F}=\left(
             \begin{array}{cccc}
               \frac{1}{[n_1/2]!} & \frac{1}{([n_1/2]-1)!} &  \frac{1}{([n_1/2]-2)!} & \cdots \\
               \frac{1}{[n_2/2]!} & \frac{1}{([n_2/2]-1)!} &  \frac{1}{([n_2/2]-2)!} & \cdots \\
                \vdots &  \vdots & \vdots & \vdots \\
                \frac{1}{[n_{k_{odd}}/2]!} & \frac{1}{([n_{k_{odd}}/2]-1)!} & \frac{1}{([n_{k_{odd}}/2]-2)!} & \cdots
             \end{array}
           \right)_{k_{odd}\times k_{odd}},   \label{defF2} \\
&& \mathbf{G} \equiv\mathbf{F}^{-1} \mathbf{D}_1 \mathbf{F}, \label{defG2} \\
&&\mathbf{R}=\left(
             \begin{array}{cccc}
               \frac{1}{[\hat{n}_1/2]!} &  \frac{1}{([\hat{n}_1/2]-1)!} &  \frac{1}{([\hat{n}_1/2]-2)!} & \cdots \\
               \frac{1}{[\hat{n}_2/2]!} &  \frac{1}{([\hat{n}_2/2]-1)!} &  \frac{1}{([\hat{n}_2/2]-2)!} & \cdots \\
                \vdots &  \vdots & \vdots & \vdots \\
               \frac{1}{[\hat{n}_{k_{even}}/2]!} &  \frac{1}{([\hat{n}_{k_{even}}/2]-1)!} & \frac{1}{([\hat{n}_{k_{even}}/2]-2)!} & \cdots
             \end{array}
           \right)_{k_{even}\times k_{odd}}. \label{defR2}
\end{eqnarray}
Here, $1/k!\equiv 0$ if $k<0$. Note that the diagonal of the $\mathbf{D}_1$ matrix is the vector of $a_{i,1}$ parameters for the odd indices of $\Lambda$, and the $\mathbf{R}$ matrix has more columns than rows since $k_{odd}>k_{even}$.

We also split matrices $\mathbf{G}$ and $\mathbf{R}$ as
\[ \label{defFRsplit}
\mathbf{G}=\left[ \mathbf{G}_1, \mathbf{G}_2\right], \quad
\mathbf{R}=\left[ \mathbf{R}_1, \mathbf{R}_2\right],
\]
where $\mathbf{G}_1$, $\mathbf{R}_1$ each have $k_{even}$ columns, while $\mathbf{G}_2$, $\mathbf{R}_2$ each have $d$ columns. Notice that $\mathbf{R}_1$ is a square $k_{even}\times k_{even}$ matrix, while $\mathbf{G}_1$ has more rows ($k_{odd}$ rows) than columns. We further define
\[ \label{defGhat}
\widehat{\mathbf{G}}\equiv \mathbf{G}_2-\mathbf{G}_1\mathbf{R}_1^{-1} \mathbf{R}_2,
\]
which is a $k_{odd}\times d$ matrix. In addition, we introduce the following $[d/2]$ minors of the $\widehat{\mathbf{G}}$ matrix,
\[\label{defMrhat0}
\widehat{M}_r\equiv \det_{k_{odd}+1-r\le i\le k_{odd}, \ 1\le j\le r}\widehat{\mathbf{G}}, \quad r=1, 2, \cdots, [d/2].
\]
These minors are determinants of $r\times r$ submatrices in the lower left corner of matrix $\widehat{\mathbf{G}}$.

\subsubsection{Notations for the case of $k_{even}-k_{odd}>2$} \label{seccase2}
In this case, there are more even indices than odd indices in $\Lambda$, and $d=k_{even}-k_{odd}-1$, where $d$ is as defined in Eq.~(\ref{defd}). Now, we group such indices such that $n_1<n_2<\cdots<n_{k_{even}}$ are all even indices, followed by $\hat{n}_{1}<\hat{n}_{2}< \cdots< \hat{n}_{k_{odd}}$ which are all odd indices, with $k_{even}+k_{odd}=N$.
That is, $\Lambda=(n_1, n_2, \dots, n_{k_{even}},\hat{n}_{1}, \hat{n}_{2}, \dots, \hat{n}_{k_{odd}})$.

In this case, we redefine matrices
\begin{eqnarray}
&& \mathbf{D}_1= \mbox{diag}\left(a_{1,1}, a_{2,1}, \cdots, a_{k_{even},1} \right),  \label{defD1b} \\
&& \mathbf{F}=\left(
             \begin{array}{cccc}
               \frac{1}{[n_1/2]!} & \frac{1}{([n_1/2]-1)!} &  \frac{1}{([n_1/2]-2)!} & \cdots \\
               \frac{1}{[n_2/2]!} & \frac{1}{([n_2/2]-1)!} &  \frac{1}{([n_2/2]-2)!} & \cdots \\
                \vdots &  \vdots & \vdots & \vdots \\
                \frac{1}{[n_{k_{even}}/2]!} & \frac{1}{([n_{k_{even}}/2]-1)!} & \frac{1}{([n_{k_{even}}/2]-2)!} & \cdots
             \end{array}
           \right)_{k_{even}\times k_{even}},   \label{defF3} \\
&& \mathbf{G} \equiv\mathbf{F}^{-1} \mathbf{D}_1 \mathbf{F}, \label{defG3} \\
&&\mathbf{R}=\left(
             \begin{array}{cccc}
               \frac{1}{[\hat{n}_1/2]!} &  \frac{1}{([\hat{n}_1/2]-1)!} &  \frac{1}{([\hat{n}_1/2]-2)!} & \cdots \\
               \frac{1}{[\hat{n}_2/2]!} &  \frac{1}{([\hat{n}_2/2]-1)!} &  \frac{1}{([\hat{n}_2/2]-2)!} & \cdots \\
                \vdots &  \vdots & \vdots & \vdots \\
               \frac{1}{[\hat{n}_{k_{odd}}/2]!} &  \frac{1}{([\hat{n}_{k_{odd}}/2]-1)!} & \frac{1}{([\hat{n}_{k_{odd}}/2]-2)!} & \cdots
             \end{array}
           \right)_{k_{odd}\times k_{even}}. \label{defR3}
\end{eqnarray}
We also split matrices $\mathbf{G}$ and $\mathbf{R}$ differently from (\ref{defFRsplit}) as
\[ \label{defFRsplitb}
\mathbf{G}=\left[ \mathbf{G}_0, \mathbf{G}_1, \mathbf{G}_2\right], \quad
\mathbf{R}=\left[ \mathbf{R}_1, \mathbf{R}_2, \mathbf{R}_3 \right],
\]
where $\mathbf{G}_0$ is $\mathbf{G}$'s first column, $\mathbf{G}_1$ its next $k_{odd}$ columns, $\mathbf{G}_2$ its remaining $d$ columns, $\mathbf{R}_1$ is $\mathbf{R}$'s first $k_{odd}$ columns, $\mathbf{R}_2$ its next $d$ columns, and $\mathbf{R}_3$ its last column. We further define
\[ \label{defGhat2}
\widehat{\mathbf{G}}\equiv \mathbf{G}_2-\mathbf{G}_1\mathbf{R}_1^{-1} \mathbf{R}_2,
\]
which is now a $k_{even}\times d$ matrix. In addition, we introduce the following $[d/2]$ minors of this $\widehat{\mathbf{G}}$ matrix,
\[\label{defMrhat2}
\widehat{M}_r\equiv \det_{k_{even}+1-r\le i\le k_{even}, \ 1\le j\le r}\widehat{\mathbf{G}}, \quad r=1, 2, \cdots, [d/2].
\]
These minors are determinants of $r\times r$ submatrices in the lower left corner of matrix $\widehat{\mathbf{G}}$.

\subsubsection{The assumption and $\hat{\beta}_{r,r}$ values}
Under the above notations, our assumption for Theorem~\ref{Theorem2} is the following.
\begin{quote}
\textbf{Assumption 2.} \emph{We assume that}
\[
\widehat{M}_r\ne 0, \quad r=1, 2, \dots, [d/2]-1; \quad \frac{\widehat{M}_{[d/2]}}{\widehat{M}_{[d/2]-1}}\ne
\left\{\begin{array}{ll}
0,  & \mbox{when $d$ is odd}, \\
-\frac{4}{3}, &  \mbox{when $d$ is even}, \end{array}\right.
\]
\emph{where $\widehat{M}_r$ is defined in Eq.~(\ref{defMrhat0}) for the case of $k_{odd}-k_{even}>1$ and in Eq.~(\ref{defMrhat2}) for the case of $k_{even}-k_{odd}>2$.}
\end{quote}

This assumption holds for generic values of $\left(a_{1,1}, a_{2,1}, a_{3,1},\cdots\right)$. But it does not hold in the previous case studied in \cite{YangYangKPI} where all $a_{i,1}$'s were equal to each other. In that case, $\mathbf{G}$ is proportional to an identity matrix and thus $\widehat{M}_1=0$, violating this assumption. Because of that, the concentric-ring pattern of lumps in the inner region as predicted by Theorem~\ref{Theorem2} does not apply to the previous case of \cite{YangYangKPI} (indeed the inner pattern in \cite{YangYangKPI} was a triangular shape of lumps instead).

Under Assumption 2, the $[d/2]\times [d/2]$ submatrix in the lower left corner of $\widehat{\mathbf{G}}$ defined in Eq.~(\ref{defGhat}) or (\ref{defGhat2}) would admit the following factorization
\[ \label{FacAB12}
\widehat{\mathbf{G}}_{\hat{k}+1-[d/2]\le i\le \hat{k}, \ 1\le j\le [d/2]}= \widehat{\mathbf{A}}\widehat{\mathbf{B}},
\]
where $\hat{k}$ is equal to $k_{odd}$ when $k_{odd}-k_{even}>1$ and equal to $k_{even}$ when $k_{even}-k_{odd}>2$,
\[ \label{formABhat}
\widehat{\mathbf{A}}=\left(
\begin{array}{ccccc}
 1 & \hat{\alpha} _{1,2} & \cdots & \hat{\alpha} _{1,[d/2]}  \\
 0 & 1 & \cdots & \hat{\alpha} _{2,[d/2]}  \\
 \vdots & \vdots & \ddots & \vdots \\
 0 & 0 & 0 & 1  \\
\end{array}
\right),\quad \widehat{\mathbf{B}} =\left(
\begin{array}{ccccc}
 0 & 0 & \cdots & \hat{\beta} _{[d/2],[d/2]}  \\
 \vdots & \vdots & .\cdot^{\cdot} & \vdots \\
 0 & \hat{\beta} _{2,2} & \cdots & \hat{\beta} _{2,[d/2]}  \\
 \hat{\beta} _{1,1} & \hat{\beta} _{1,2} & \cdots & \hat{\beta} _{1,[d/2]} \\
\end{array}
\right),
\]
and $\hat{\alpha}_{i,j}$, $\hat{\beta}_{i,j}$ are complex constants. In particular,
\[ \label{formbetahat}
\hat{\beta} _{1,1}=\widehat{M}_1, \quad \hat{\beta}_{r,r}=\frac{\widehat{M}_{r}}{\widehat{M}_{r-1}}, \quad 1<r\le [d/2].
\]
Under Assumption 2, $\hat{\beta}_{r,r}\ne 0$ for $1\le r\le [d/2]-1$, and $\hat{\beta}_{[d/2], [d/2]}\ne 0$ when $d$ is odd and
$\hat{\beta}_{[d/2], [d/2]}+4/3\ne 0$ when $d$ is even. These $\hat{\beta}_{r,r}$ values are the ones mentioned in Theorem~\ref{Theorem2}. One can see that this $\widehat{\mathbf{A}}\widehat{\mathbf{B}}$ factorization is the counterpart of a similar factorization in Eq.~(\ref{FacAB}) for the $\Lambda=(1, 3, \dots, 2N-1)$ case.

\subsection{Remarks on the center-lump location for odd values of $d$} \label{secRemarks}
Theorem~\ref{Theorem2} predicts that when $d$ is odd, there would also be a fundamental lump lying in the $O(1)$ region of the ring center $(x, y)=(12t, 0)$. The asymptotic location of this center lump was not provided in Theorem~\ref{Theorem2} because this is a minor point and we do not want to spend much space on it. To derive its asymptotic position, one only needs to slightly modify our calculations leading to the asymptotic positions of fundamental lumps on concentric rings. We will only mention the results below without providing details.

When $d>1$, we take the bottom left $([d/2]+1)\times ([d/2]+1)$ submatrix of $\widehat{\mathbf{G}}$ in Eq.~(\ref{defGhat}) or (\ref{defGhat2}), and increase its two matrix elements adjacent to its top-right corner by $4/3$ and call this new matrix $\widetilde{\mathbf{G}}$. Then, we perform the $\widetilde{\mathbf{A}}\widetilde{\mathbf{B}}$ factorization to this new matrix $\widetilde{\mathbf{G}}$ similar to Eq.~(\ref{FacAB12}). Then, the leading-order position $(x_0, y_0)$ of the center lump would be predicted by $x_0+2{\rm{i}}y_0=-\tilde{\beta}_{[d/2]+1, [d/2]+1}$, where $\tilde{\beta}_{[d/2]+1, [d/2]+1}$ is from the $\widetilde{\mathbf{B}}$ matrix of that factorization, and the error of this $(x_0, y_0)$ prediction is $O(|t|^{-1}$). Notice that this result is similar to the $\Lambda=(1, 3, \dots, 2N-1)$ case briefly described at the end of  Sec.~\ref{secMainResult}.

When $d=1$, the situation is a little different. In this case, we need to extend the $\mathbf{F}$ matrix in Eq.~(\ref{defF2}) or (\ref{defF3}) by one column and call this new column $\mathbf{F}_2$. We also define a new vector $\mathbf{W}=\mathbf{F}^{-1}\mathbf{F}_2$, where $\mathbf{F}$ is the pre-extended square matrix. Then, the leading-order position $(x_0, y_0)$ of the center lump would be predicted by $x_0+2{\rm{i}}y_0=-\left(\widehat{M}_1+(4/3)w_{last}\right)$, where $\widehat{M}_1$ is as defined in Eq.~(\ref{defMrhat0}) or (\ref{defMrhat2}), and $w_{last}$ is the last element of the vector $\mathbf{W}$. The error of this $(x_0, y_0)$ prediction is $O(|t|^{-1}$).

\subsection{Numerical verifications of Theorem 2}
Now we use two examples to verify Theorem~\ref{Theorem2}.

\begin{figure}[htbp]
\begin{center}
\vspace{8.0cm}
\hspace{-2.0cm}
\includegraphics[scale=0.25, bb=300 0 400 360]{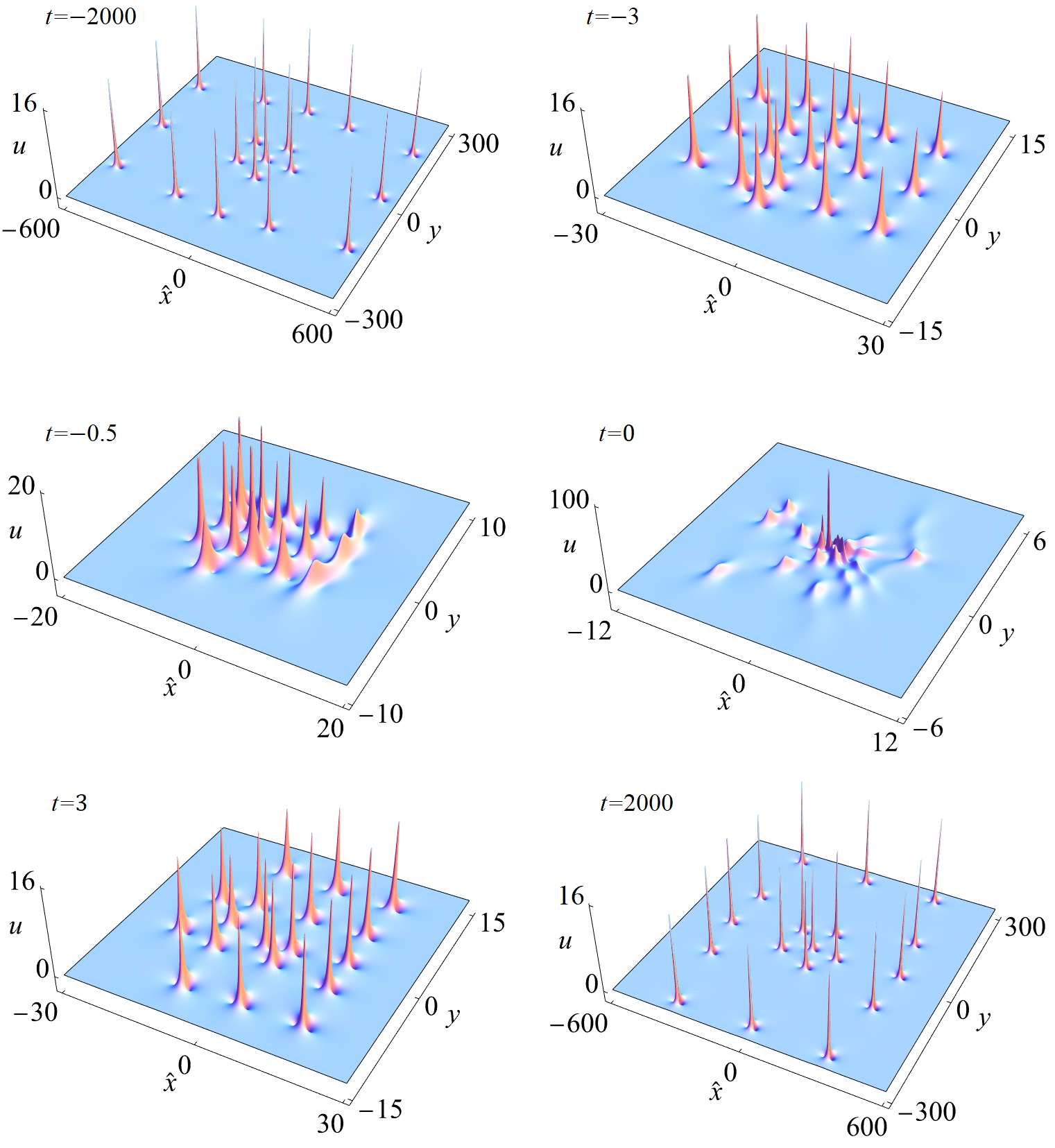}\hspace{2.0cm}
\caption{The true solution $u_{\Lambda} (x,y,t)$ with $\Lambda =(3, 5, 7, 9, 4)$ at time values of $t=-2000, -3, -0.5, 0, 3$ and 2000. Internal parameters are taken as in Eq.~(\ref{para3}), with the other elements of parameter vectors as zero. The axis $\hat{x}=x-12t$ is the moving $x$ coordinate. } \label{f:fig6}
 \end{center}
\end{figure}

\textbf{Example 3.} In this example, we take $N=5$ with $\Lambda =(3, 5, 7, 9, 4)$, which falls into the case of $k_{odd}-k_{even}>1$. Internal parameters $\textbf{\emph{a}}_1, \dots, \textbf{\emph{a}}_5$ are taken as
\[ \label{para3}
(a_{1,1}, a_{2,1}, a_{3,1}, a_{4,1}, a_{5,1})=(0, {\rm{i}}, 2, 2{\rm{i}}, -1),
\]
with the other elements of parameter vectors $\textbf{\emph{a}}_i \ (1\le i\le 5)$ taken as zero. The true solution from Lemma~1 at six time values of $t=-2000, -3, -0.5, 0, 3$ and 2000 is plotted in Fig.~\ref{f:fig6}. It is seen that at $t=-2000$, the solution splits into eighteen fundamental lumps. Twelve of them are located in the outer region, forming a quasi-rectangular shape. The other six are located in the inner region, with five of them lying on a ring and the remaining lump near the ring center. As time increases to $-3$, $-0.5$ and 0, these eighteen fundamental lumps move close to each other and coalesce, forming a high spike plus some low wave structures. As time increases to $3$, the coalesced solution splits up into eighteen fundamental lumps again. At $t=2000$, these eighteen fundamental lumps evolve into a clear pattern, with twelve of them located in the outer region forming a quasi-rectangular shape that is rotated $90^\circ$ from that in the $t=-2000$ solution. The remaining six fundamental lumps are located in the inner region forming a ring of five fundamental lumps plus another lump near the ring center, similar to that in the $t=-2000$ solution. In addition, the relative positions of fundamental lumps on the ring in the inner region at $t=\pm 2000$ are roughly the same.

\begin{figure}[htb]
\begin{center}
\vspace{-0.0cm}
\includegraphics[scale=0.25, bb=0 0 1150 490]{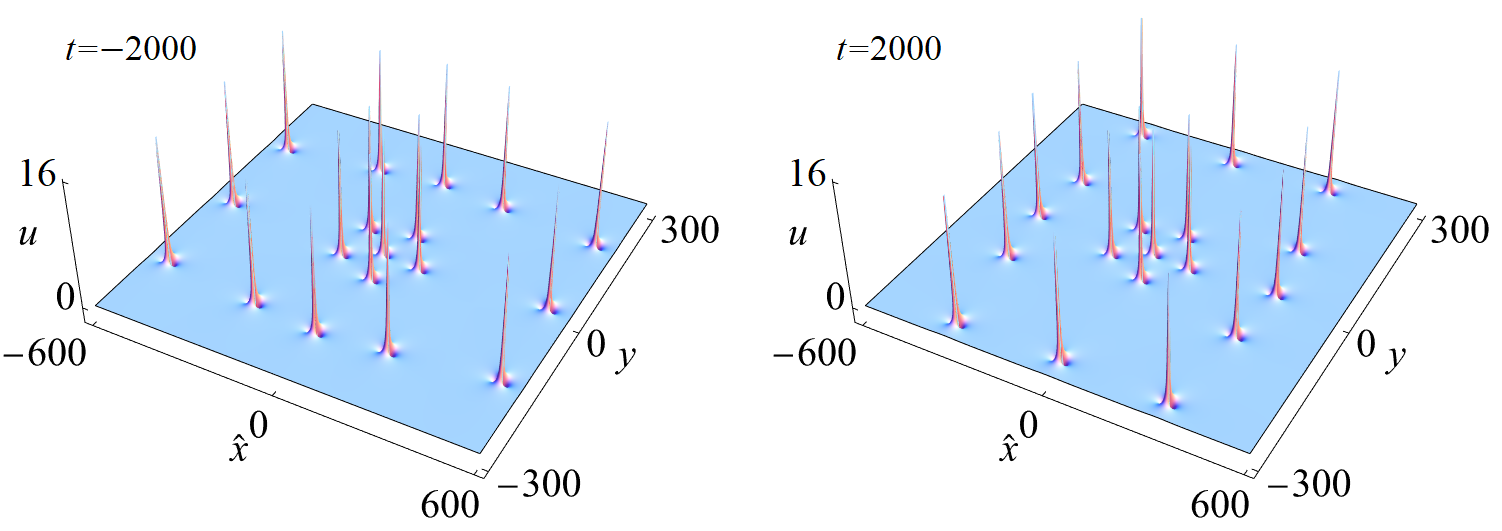}
\caption{Leading-order predictions of lump patterns from Theorem~\ref{Theorem2} for the solution of Fig.~\ref{f:fig6} at large times of $t=\pm 2000$. }  \label{f:fig7}
\end{center}
\end{figure}

Now, we use Theorem~\ref{Theorem2} to analytically predict the solution of Fig.~\ref{f:fig6} at large times of $t=\pm 2000$. In this case, $k_{odd}=4$ and $k_{even}=1$. Thus, this belongs to the case of $k_{odd}-k_{even}>1$ with $d=3$. The Wronskian-Hermit polynomial $W_{\Lambda}(z)$ for $\Lambda =(3, 5, 7, 9, 4)$ has 12 nonzero simple roots, which are approximately $\pm 3.4177$, $\pm 2.5684\textrm{i}$, $\pm 1.3866 \pm 2.4091 \textrm{i}$  and $\pm 3.4609 \pm 2.1538 \textrm{i}$. Theorem~\ref{Theorem2} then predicts that the solution of Fig.~\ref{f:fig6} at large times would split into twelve fundamental lumps in the outer region, one ring with five fundamental lumps on it in the inner region, plus another fundamental lump near the ring center. Locations of those fundamental lumps in the outer region are predicted by Eq.~(\ref{x0t0r8}), with $z_0$ being the above twelve roots. To predict locations of fundamental lumps on the ring, we notice from the parameter choices (\ref{para3}) and $\mathbf{D}_1$'s definition (\ref{defD1}) that $\mathbf{D}_1= \mbox{diag}(0, {\rm{i}}, 2, 2{\rm{i}})$. Thus, the $\mathbf{G}$ matrix from Eq.~(\ref{defG2}) is
\[
\mathbf{G}=\left(
         \begin{array}{cccc}
           1 & 1 & 0 & 0 \\
           \frac{1}{2!} & 1 & 1 & 0 \\
           \frac{1}{3!} & \frac{1}{2!} & 1 & 1 \\
         \frac{1}{4!} & \frac{1}{3!} & \frac{1}{2!} & 1 \\
         \end{array}
       \right)^{-1}
\left(
  \begin{array}{cccc}
    0 & 0 & 0 & 0 \\
    0 & \textrm{i} & 0 & 0  \\
    0 & 0 & 2 & 0  \\
    0 & 0 & 0 & 2 \textrm{i}  \\
  \end{array}
\right)
\left(
         \begin{array}{cccc}
           1 & 1 & 0 & 0 \\
           \frac{1}{2!} & 1 & 1 & 0 \\
           \frac{1}{3!} & \frac{1}{2!} & 1 & 1 \\
         \frac{1}{4!} & \frac{1}{3!} & \frac{1}{2!} & 1 \\
         \end{array}
       \right)=
\left(
\begin{array}{cccc}
 8-8\textrm{i} & 24-20\textrm{i} & 48-36\textrm{i} & 48-48\textrm{i} \\
 -8+8\textrm{i} & -24+20\textrm{i} & -48+36\textrm{i} & -48+48\textrm{i} \\
 4-\frac{7\textrm{i}}{2} & 12-9\textrm{i} & 24-17\textrm{i} & 24-24\textrm{i} \\
 -1+\frac{5\textrm{i}}{6} & -3+\frac{7\textrm{i}}{3} & -6+5\textrm{i} & -6+8\textrm{i} \\
\end{array}
\right),
\]
and the $\mathbf{R}$ matrix from Eq.~(\ref{defR2}) is
\[
\mathbf{R}=\left(1/2, 1, 1, 0\right).
\]
Then, we can calculate the $\widehat{\mathbf{G}}$ matrix from Eq.~(\ref{defGhat}) and get $\widehat{M}_1=\hat{\beta}_{1,1}= -1+\frac{2}{3}\textrm{i}$. Using this $\hat{\beta}_{1,1}$ value, we can obtain leading-order predictions of lump positions on this ring from Eq.~(\ref{x0t0r9}) with $d=3$ and $r=1$. These predicted outer-region and inner-region-ring solutions at large times of $t=\pm 2000$ are plotted in Fig.~\ref{f:fig7} (the center lump whose position is predicted by the remarks of Sec.~\ref{secRemarks} is also shown for completeness). Comparing these predictions with true solutions at $t=\pm 2000$ in Fig.~\ref{f:fig6}, we can see that the predictions agree with true solutions very well.

\begin{figure}[htb]
\begin{center}
\includegraphics[scale=0.35, bb=-300 0 1150 400]{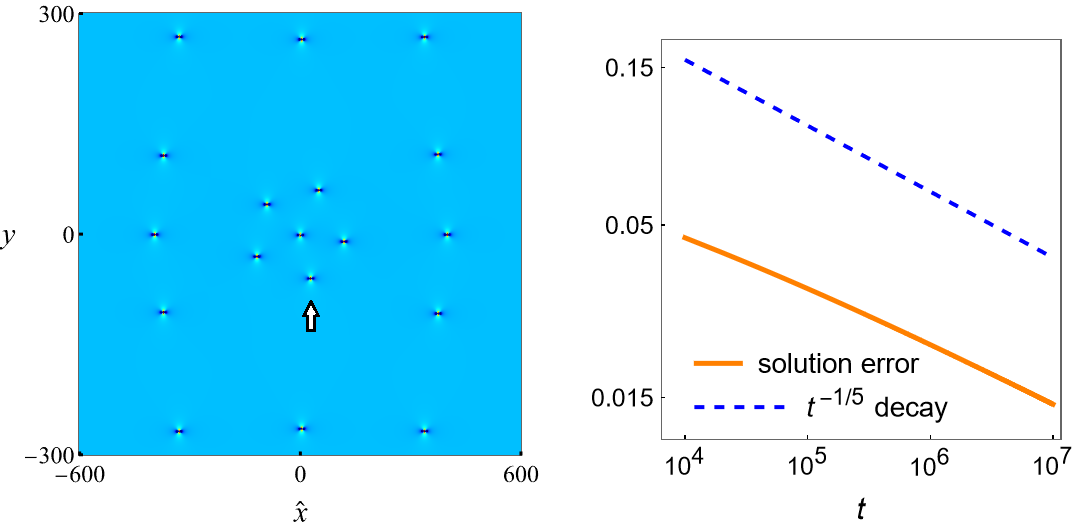}
\caption{Verification of the decay rate of relative error in leading-order predictions of fundamental lumps' positions in Eq.~(\ref{x0t0r9}) of Theorem~\ref{Theorem2} for the example of Fig.~\ref{f:fig6} with $\Lambda=(3, 5, 7, 9, 4)$. The left panel shows the density plot of the higher-order lump solution in Fig.~\ref{f:fig6} at $t=2000$. The right panel shows the relative error versus time $t$ for the location of the lump on the ring marked by a vertical white arrow in the left panel (the predicted $|t|^{-1/5}$ decay is also plotted as a dashed line for comparison).  }  \label{f:fig8}
\end{center}
\end{figure}

Quantitative comparisons between the predicted and true solutions in Example 3 at large times have also been made in order to verify the decay rate of relative errors on fundamental lumps' positions (\ref{x0t0r9}) of the inner ring. For this purpose, the density plot of the true higher-order lump solution in Fig.~\ref{f:fig6} at $t=2000$ is displayed in the left panel of Fig.~\ref{f:fig8}. We then pick a fundamental lump on the inner ring that is marked by a vertical white arrow in that panel. For this fundamental lump, we numerically determine at each large time $t$ the relative error of prediction (\ref{x0t0r9}) for its position. The graph of this relative error versus time $t$ is plotted in the right panel of Fig.~\ref{f:fig8}. The predicted relative error from Eq.~(\ref{x0t0r9}) (with $d=3$ and $r=1$) is $O(|t|^{-1/5})$. This predicted decay rate is also plotted as a dashed line in the right panel for comparison. We can see from this panel that the true decay rate indeed agrees with the prediction at large time, which quantitatively confirms Theorem~\ref{Theorem2}.

\begin{figure}[htbp]
\begin{center}
\vspace{8.0cm}
\hspace{-2.0cm}
\includegraphics[scale=0.25, bb=300 0 400 300]{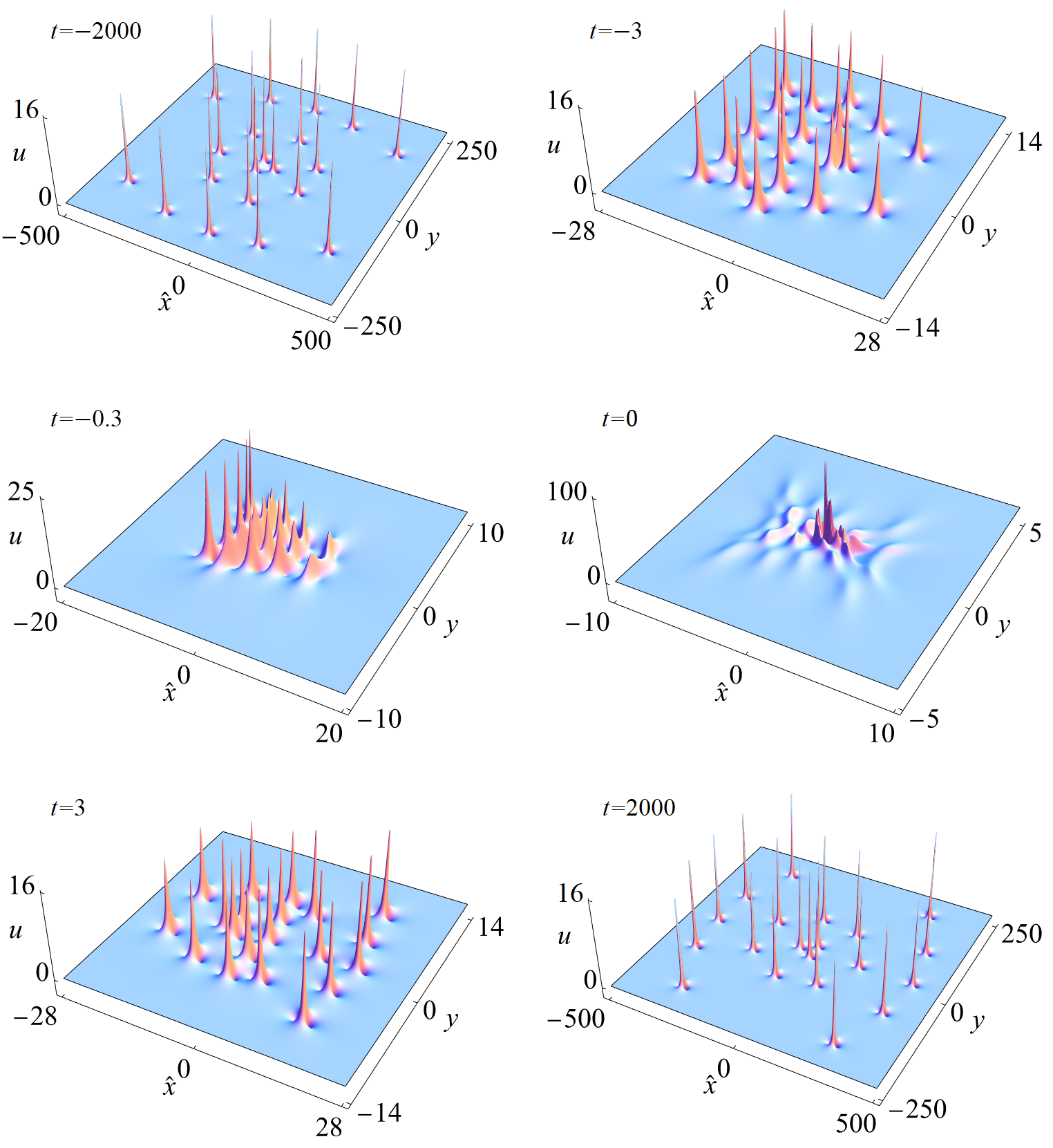}\hspace{2.0cm}
\caption{The true solution $u_{\Lambda} (x,y,t)$ with $\Lambda =(2, 4, 6, 8, 10)$ at time values of $t=-2000,  -3, =0.3, 0, 3$ and 2000. Internal parameters are taken as in Eq.~(\ref{para4}), with the other elements of parameter vectors as zero. The axis $\hat{x}=x-12t$ is the moving $x$ coordinate. } \label{f:fig9}
 \end{center}
\end{figure}

\textbf{Example 4.}
As the last example, we take $N=5$ with $\Lambda =(2, 4, 6, 8, 10)$, which falls into the case of $k_{even}-k_{odd}>2$. Internal parameters $\textbf{\emph{a}}_1, \dots, \textbf{\emph{a}}_5$ are taken as
\[ \label{para4}
(a_{1,1}, a_{2,1}, a_{3,1}, a_{4,1}, a_{5,1})=(0, 1, 1, 1, -1),
\]
with the other elements of parameter vectors $\textbf{\emph{a}}_i \ (1\le i\le 5)$ taken as zero. The true solution from Lemma~1 at six time values of $t=-2000,  -3, -0.3, 0, 3$ and 2000 is plotted in Fig.~\ref{f:fig9}.
It is seen that at $t=-2000$, the solution splits into twenty fundamental lumps. Ten of them are located in the outer region, forming two disjoint arc segments. The other ten are located on two concentric rings of 7 and 3 fundamental lumps each in the inner region. As time increases to $-3$, $-0.3$ and 0, these twenty fundamental lumps move close to each other and coalesce, forming a high spike and some low wave structures. As time increases to $3$, the coalesced solution splits up into twenty fundamental lumps again. At $t=2000$, these twenty fundamental lumps evolve into a clear pattern, with ten of them located in the outer region forming two disjoint arcs that are rotated $90^\circ$ from that in the $t=-2000$ solution. The remaining ten fundamental lumps are located on two concentric rings of 7 and 3 fundamental lumps each in the inner region similar to the $t=-2000$ solution, but the relative positions of fundamental lumps on these two rings in the inner region are different at $t=-2000$ and $2000$.

\begin{figure}[htb]
\begin{center}
\vspace{-0.0cm}
\includegraphics[scale=0.25, bb=0 0 1150 490]{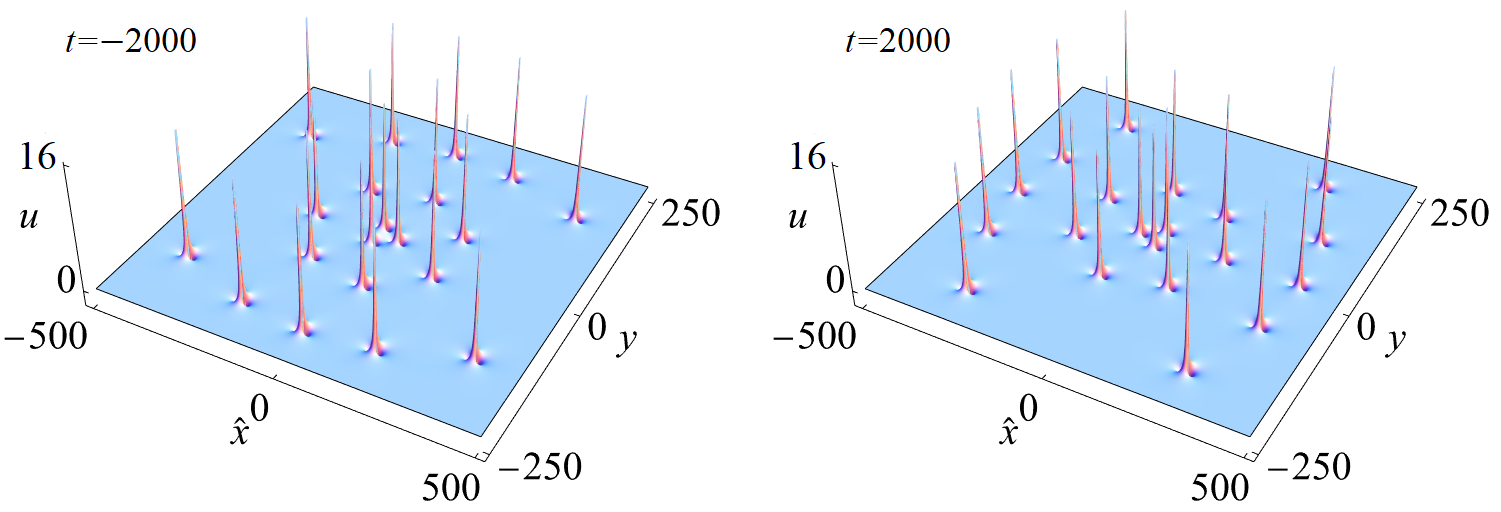}
\caption{Leading-order predictions of lump patterns from Theorem~\ref{Theorem2} for the solution of Fig.~\ref{f:fig9} at large times of $t=\pm 2000$. }  \label{f:fig10}
\end{center}
\end{figure}

Now, we use Theorem~\ref{Theorem2} to analytically predict the solution of Fig.~\ref{f:fig9} at large times of $t=\pm 2000$. In this case, $k_{odd}=0$ and $k_{even}=5$. Thus, this belongs to the case of $k_{even}-k_{odd}>2$ with $d=4$. The Wronskian-Hermit polynomial $W_{\Lambda}(z)$ for $\Lambda =(2,4,6,8,10)$ has ten nonzero simple roots, which are approximately $\pm  2.537203 \textrm{i} $, $\pm 2.487972 \pm 1.879695 \textrm{i} $  and $\pm 1.142561 \pm 2.396560 \textrm{i} $. Theorem~\ref{Theorem2} then predicts that the solution of Fig.~\ref{f:fig9} at large times would split into ten fundamental lumps in the outer region, and two rings with 7 and 3 fundamental lumps respectively in the inner region. Locations of those fundamental lumps in the outer region are predicted by Eq.~(\ref{x0t0r8}) with $z_0$ being the above ten roots. To predict locations of fundamental lumps on the two rings, we notice from the parameter choices (\ref{para4}) and $\mathbf{D}_1$'s definition (\ref{defD1b}) that $\mathbf{D}_1= \mbox{diag}(0, 1, 1, 1, -1)$. Thus, the $\mathbf{G}$ matrix from Eq.~(\ref{defG3}) is
\[
G=\left(
\begin{array}{ccccc}
 -6 & -15 & -40 & -120 & -240 \\
 6 & 15 & 40 & 120 & 240 \\
 -\frac{5}{2} & -\frac{13}{2} & -19 & -60 & -120 \\
 \frac{2}{3} & 2 & \frac{20}{3} & 21 & 40 \\
 -\frac{1}{8} & -\frac{11}{24} & -\frac{5}{3} & -5 & -9 \\
\end{array}
\right).
\]
Both the $\mathbf{R}$ matrix from Eq.~(\ref{defR3}) and the $\mathbf{G}_1$ matrix from Eq.~(\ref{defFRsplitb}) are null.
Thus, the $\widehat{\mathbf{G}}$ matrix in Eq.~(\ref{defGhat2}) is just the above $\mathbf{G}$ matrix with its first column removed. Then, the $2\times 2$ submatrix at the lower left corner of $\widehat{\mathbf{G}}$ and its factored $\widehat{\mathbf{B}}$ matrix from Eq.~(\ref{FacAB12}) are
\[
\widehat{\mathbf{G}}_{4\le i\le 5, \ 1\le j\le 2}=
\left(
\begin{array}{cc}
 2 & \frac{20}{3} \\
 -\frac{11}{24} & -\frac{5}{3}
\end{array}
\right), \quad \widehat{\mathbf{B}}=\left(
\begin{array}{cc}
 0 & -\frac{20}{33}   \\
 -\frac{11}{24} & -\frac{5}{3} \\
\end{array}
\right).
\]
This shows that $\beta_{1,1}=-11/24$ and $\beta_{2,2}=-20/33$. Notice that the above submatrix of $\widehat{\mathbf{G}}$ satisfies our Assumption 2. Using these values, we can obtain leading-order predictions of lump positions on these two rings from Eq.~(\ref{x0t0r9}) with $d=4$ and $r=1, 2$. These predicted solutions at large times of $t=\pm 2000$ are plotted in Fig.~\ref{f:fig10}. Comparing these predictions with true solutions at $t=\pm 2000$ in Fig.~\ref{f:fig9}, we can see that the predictions agree with true solutions very well.

\section{Proof of Theorem~\ref{Theorem2} for lump patterns with $\Lambda\ne (1, 3, \dots, 2N-1)$} \label{secProof2}

In this section, we prove Theorem~\ref{Theorem2}. Our starting point is to rewrite the determinant $\sigma$ in Eq.~(\ref{Blockmatrix}) as a larger determinant
\[ \label{3Nby3Ndet2b}
\sigma=\left|\begin{array}{cc}
\textbf{O}_{N\times N} & \Phi_{N\times \widehat{N}} \\
-\Psi_{\widehat{N}\times N} & \textbf{I}_{\widehat{N} \times \widehat{N}} \end{array}\right|,
\]
where
\[
\Phi_{i,j}=2^{-(j-1)} S_{n_i+1-j}\left(\textbf{\emph{x}}^{+} + (j-1) \textbf{\emph{s}} +\textbf{\emph{a}}_i\right), \quad \Psi_{i,j}=2^{-(i-1)} S_{n_j+1-i}\left((\textbf{\emph{x}}^{+})^* + (i-1) \textbf{\emph{s}}+\textbf{\emph{a}}_j^*\right),
\]
and $\widehat{N}=1+\max\{n_i, 1\le i\le N\}$. As we have mentioned in Sec.~\ref{secAssumptions}, we only need to consider the two cases of $k_{odd}-k_{even}>1$ and $k_{even}-k_{odd}>2$.

\subsection{Proof for the case of $k_{odd}-k_{even}>1$}
In this case, we regroup the odd indices together in the order $n_1<n_2<\cdots<n_{k_{odd}}$, followed by even indices in the order $\hat{n}_{1}<\hat{n}_{2}< \cdots< \hat{n}_{k_{even}}$, as we have done in Sec.~\ref{seccase1}. That is, $\Lambda=(n_1, \dots, n_{k_{odd}}, \hat{n}_{1}, \dots, \hat{n}_{k_{even}})$, with $k_{even}+k_{odd}=N$. As a consequence, the first $k_{odd}$ rows of $\Phi$ above correspond to odd indices, and the last $k_{even}$ rows of $\Phi$ correspond to even indices.

Let us define $T_2$-dependent matrices
\begin{eqnarray}
&& {\cal F}=\left(
             \begin{array}{cccc}
               \frac{T_2^{[n_1/2]}}{[n_1/2]!} & \frac{T_2^{[n_1/2]-1}}{([n_1/2]-1)!} & \frac{T_2^{[n_1/2]-2}}{([n_1/2]-2)!}  & \cdots \\
                \vdots &  \vdots & \vdots & \vdots \\
                \frac{T_2^{[n_{k_{odd}}/2]}}{[n_{k_{odd}}/2]!} & \frac{T_2^{[n_{k_{odd}}/2]-1}}{([n_{k_{odd}}/2]-1)!} & \frac{T_2^{[n_{k_{odd}}/2]-2}}{([n_{k_{odd}}/2]-2)!}& \cdots
             \end{array}
           \right)_{k_{odd}\times k_{odd}},   \\
&& {\cal R}=\left(
             \begin{array}{cccc}
               \frac{T_2^{[\hat{n}_1/2]}}{[\hat{n}_1/2]!} &  \frac{T_2^{[\hat{n}_1/2]-1}}{([\hat{n}_1/2]-1)!} &  \frac{T_2^{[\hat{n}_1/2]-2}}{([\hat{n}_1/2]-2)!} & \cdots \\
                \vdots &  \vdots & \vdots & \vdots \\
               \frac{T_2^{[\hat{n}_{k_{even}}/2]}}{([\hat{n}_{k_{even}}/2])!} &  \frac{T_2^{[\hat{n}_{k_{even}}/2]-1}}{([\hat{n}_{k_{even}}/2]-1)!} & \frac{T_2^{[\hat{n}_{k_{even}}/2]-2}}{([\hat{n}_{k_{even}}/2]-2)!} & \cdots
             \end{array}
 \right)_{{k_{even}}\times k_{odd}},  \\
 && \mathbf{E}=\mbox{diag}\left(1, T_2^{-1}, T_2^{-2}, \cdots T_2^{-({k_{odd}}-1)} \right).
\end{eqnarray}
Here, $T_2^k/k!\equiv0$ if $k<0$. Notice that ${\cal F}$ and ${\cal R}$ are the counterparts of $T_2$-free matrices $\mathbf{F}$ and $\mathbf{R}$ defined in Eqs. (\ref{defF2}) and (\ref{defR2}). We also split
\[ \label{defRsplit}
{\cal R}=\left[ {\cal R}_1, {\cal R}_2\right], \quad \mathbf{E}=\mbox{diag}(\mathbf{E}_1, \mathbf{E}_2),
\]
where ${\cal R}_1$ has $k_{even}$ columns, and
\[
\mathbf{E}_1=\mbox{diag}\left(1, T_2^{-1}, T_2^{-2}, \cdots T_2^{-({k_{even}}-1)} \right), \quad \mathbf{E}_2=\mbox{diag}\left(T_2^{-{k_{even}}}, T_2^{-{k_{even}}-1}, \cdots T_2^{-(k_{odd}-1)} \right)_{d\times d}.
\]
Matrices $({{\cal F}, \cal R}_1, {\cal R}_2)$ are related to $(\mathbf{F}, \mathbf{R}_1, \mathbf{R}_2)$ of Sec.~\ref{seccase1} as
\[ \label{calFR2FR}
{\cal F}=\mathbf{E}_f \mathbf{F} \mathbf{E}, \quad {\cal R}_1=\mathbf{E}_r \mathbf{R}_1 \mathbf{E}_1,\ \ \ {\cal R}_{2}=\mathbf{E}_r \mathbf{R}_2 \mathbf{E}_2,
\]
where
\[
\mathbf{E}_f=\mbox{diag}\left(T_2^{[n_1/2]}, T_2^{[n_2/2]},  \cdots T_2^{[n_{k_{odd}}/2]} \right), \quad
\mathbf{E}_r=\mbox{diag}\left(T_2^{[\hat{n}_1/2]}, T_2^{[\hat{n}_2/2]},  \cdots T_2^{[\hat{n}_{k_{even}}/2]} \right).
\]

When $1\ll |x_1^+|\ll O(|t|^{1/2})$, we use relations (\ref{polyrelation3})-(\ref{polyrelation2}) and (\ref{factS1}) to write the dominant terms of $\Phi$ in Eq.~(\ref{3Nby3Ndet2b}) as
\[ \label{PhiAasym}
\Phi_{\small{\mbox{first $k_{odd}$ rows}}}  \sim {\cal F} \mathbf{P}_1+\mathbf{D}_1{\cal F} \mathbf{P}_2,
\]
\[ \label{PhiBasym}
\Phi_{\small{\mbox{last $k_{even}$ rows}}}  \sim {\cal R} \mathbf{P}_2+\mathbf{D}_2{\cal R} \mathbf{P}_3,
\]
where $\mathbf{D}_1$ is as defined in Eq.~(\ref{defD1}), and
\begin{eqnarray}
&& \mathbf{P}_k= \mbox{Mat}_{1\leq i\leq k_{odd}, \ 1\le j\le \widehat{N}} \left( 2^{-(j-1)} S_{2i-j+1-k}(\textbf{v})\right), \quad k=1, 2, 3,  \label{defPk123} \\
&& \mathbf{D}_2= \mbox{diag}\left(a_{k_{odd}+1,1}, a_{k_{odd}+2,1}, \cdots, a_{N,1} \right).
\end{eqnarray}
Notice that the diagonal of matrix $\mathbf{D}_2$ is the vector of $a_{i,1}$ parameters for the even indices of $\Lambda$. We also split the matrix $\mathbf{P}_2$ as
\[
\mathbf{P}_2=\left(\begin{array}{c} \mathbf{P}_{2}^{(u)} \\ \mathbf{P}_{2}^{(l)}\end{array}\right),
\]
where $\mathbf{P}_{2}^{(u)}$ is the first $k_{even}$ rows of $\mathbf{P}_2$. Then, employing Eqs.~(\ref{defG2}), (\ref{defFRsplit}) and (\ref{calFR2FR}), we can rewrite the asymptotics (\ref{PhiAasym}) and (\ref{PhiBasym}) of $\Phi$ as
\[ \label{PhiAasym2}
\Phi_{\small{\mbox{first $k_{odd}$ rows}}}  \sim {\cal F} \left(\mathbf{P}_1+\mathbf{E}^{-1}\mathbf{G} \mathbf{E}\mathbf{P}_2\right)={\cal F} \left(\mathbf{P}_1 + \mathbf{E}^{-1} \mathbf{G}_1\mathbf{E}_1\mathbf{P}_{2}^{(u)}+ \mathbf{E}^{-1} \mathbf{G}_2\mathbf{E}_2\mathbf{P}_{2}^{(l)}\right),
\]
and
\[ \label{PhiBasym2}
\Phi_{\small{\mbox{last $k_{even}$ rows}}} \sim
{\cal R}_1\mathbf{P}_{2}^{(u)}+{\cal R}_{2}\mathbf{P}_{2}^{(l)} +\mathbf{D}_2{\cal R} \mathbf{P}_3={\cal R}_1\left\{
\mathbf{P}_{2}^{(u)}+ \mathbf{E}_1^{-1}\mathbf{R}_1^{-1} \mathbf{R}_2\mathbf{E}_2\mathbf{P}_{2}^{(l)}+\mathbf{E}_1^{-1}\mathbf{R}_1^{-1}\mathbf{D}_2\mathbf{R}\mathbf{E} \mathbf{P}_3\right\}.
\]
Thus,
\begin{eqnarray}
\Phi\sim \left( \begin{array}{cc}
                  {\cal F} & \textbf{O} \\
                  \textbf{O} & {\cal R}_1
                \end{array}\right)
\left(\begin{array}{l}
 \mathbf{P}_1 + \mathbf{E}^{-1} \mathbf{G}_1\mathbf{E}_1\mathbf{P}_{2}^{(u)}+ \mathbf{E}^{-1} \mathbf{G}_2\mathbf{E}_2\mathbf{P}_{2}^{(l)} \\
 \mathbf{P}_{2}^{(u)}+\mathbf{E}_1^{-1}\mathbf{R}_1^{-1} \mathbf{R}_2\mathbf{E}_2\mathbf{P}_{2}^{(l)}
 +\mathbf{E}_1^{-1}\mathbf{R}_1^{-1}\mathbf{D}_2\mathbf{R}\mathbf{E} \mathbf{P}_3
\end{array}\right).
\end{eqnarray}

The key step of this proof is to use the lower $\mathbf{P}_{2}^{(u)}$ block of the above right matrix to eliminate the upper $\mathbf{E}^{-1} \mathbf{G}_1\mathbf{E}_1\mathbf{P}_{2}^{(u)}$ term of that matrix through row operations, which does not affect the $\sigma$ determinant in Eq.~(\ref{3Nby3Ndet2b}). From this step we get
\[ \label{detPhi20}
\Phi\sim \left( \begin{array}{cc}
                  {\cal F} & \textbf{O} \\
                  \textbf{O} & {\cal R}_1
                \end{array}\right) \left(\begin{array}{l}
 \mathbf{P}_1 + \mathbf{E}^{-1} \widehat{\mathbf{G}}\mathbf{E}_2\mathbf{P}_{2}^{(l)}-
 \mathbf{E}^{-1}\mathbf{G}_1\mathbf{R}_1^{-1}\mathbf{D}_2\mathbf{R}\mathbf{E}\mathbf{P}_3 \\
\mathbf{P}_{2}^{(u)}+\mathbf{E}_1^{-1}\mathbf{R}_1^{-1} \mathbf{R}_2\mathbf{E}_2\mathbf{P}_{2}^{(l)}
 +\mathbf{E}_1^{-1}\mathbf{R}_1^{-1}\mathbf{D}_2\mathbf{R}\mathbf{E} \mathbf{P}_3
\end{array}\right),
\]
where $\widehat{\mathbf{G}}$ is as defined in Eq.~(\ref{defGhat}).

Notice that the first row of matrix $\mathbf{P}_3$ is all zero, and its later $i$-th row $(i>1)$ is just the $(i-1)$-th row of $\mathbf{P}_1$. As such, when $|x_1^+|\gg 1$, it is easy to see that the term $\mathbf{E}^{-1}\mathbf{G}_1\mathbf{R}_1^{-1}\mathbf{D}_2\mathbf{R}\mathbf{E}\mathbf{P}_3$ in the upper row block of the above right matrix is subdominant to $\mathbf{P}_1$ and can be asymptotically neglected at large $|t|$. Thus, Eq.~(\ref{detPhi20}) reduces to
\begin{eqnarray} \label{detPhi12}
\Phi\sim \left( \begin{array}{cc}
                  {\cal F} & \textbf{O} \\
                  \textbf{O} & {\cal R}_1
                \end{array}\right) \left(\begin{array}{l}
 \mathbf{P}_1 + \mathbf{E}^{-1} \widehat{\mathbf{G}}\mathbf{E}_2\mathbf{P}_{2}^{(l)} \\
\mathbf{P}_{2}^{(u)}+\mathbf{E}_1^{-1}\mathbf{R}_1^{-1} \mathbf{R}_2\mathbf{E}_2\mathbf{P}_{2}^{(l)}
 +\mathbf{E}_1^{-1}\mathbf{R}_1^{-1}\mathbf{D}_2\mathbf{R}\mathbf{E} \mathbf{P}_3
\end{array}\right).
\end{eqnarray}
We also use $\mathbf{P}_1$ in the first row block of the above right matrix to eliminate the $\mathbf{P}_3$ term in the second row block of the above right matrix in view of the simple relation between $\mathbf{P}_1$ and $\mathbf{P}_3$ mentioned above. This operation would introduce only subdominant terms to the existing $\mathbf{E}_1^{-1}\mathbf{R}_1^{-1} \mathbf{R}_2\mathbf{E}_2\mathbf{P}_{2}^{(l)}$ term in the second row block as well when $|x_1^+|\gg 1$. Thus, we can asymptotically neglect that $\mathbf{P}_3$ term in the second row block and reduce (\ref{detPhi12}) to
\begin{eqnarray} \label{detPhi13}
\Phi\sim \left( \begin{array}{cc}
                  {\cal F} & \textbf{O} \\
                  \textbf{O} & {\cal R}_1
                \end{array}\right) \left(\begin{array}{l}
 \mathbf{P}_1 + \mathbf{E}^{-1} \widehat{\mathbf{G}}\mathbf{E}_2\mathbf{P}_{2}^{(l)} \\
\mathbf{P}_{2}^{(u)}+\mathbf{E}_1^{-1}\mathbf{R}_1^{-1} \mathbf{R}_2\mathbf{E}_2\mathbf{P}_{2}^{(l)}
\end{array}\right).
\end{eqnarray}
To proceed further, we see from Eq.~(\ref{factS1}) that the $\mathbf{E}_1^{-1}\mathbf{R}_1^{-1} \mathbf{R}_2\mathbf{E}_2\mathbf{P}_{2}^{(l)}$ term in the above right matrix is subdominant to its $\mathbf{P}_{2}^{(u)}$ term, which reduces Eq.~(\ref{detPhi13}) to
\begin{eqnarray} \label{detPhi14}
\Phi\sim \left( \begin{array}{cc}
                  {\cal F} & \textbf{O} \\
                  \textbf{O} & {\cal R}_1
                \end{array}\right) \left(\begin{array}{l}
 \mathbf{P}_1 + \mathbf{E}^{-1} \widehat{\mathbf{G}}\mathbf{E}_2\mathbf{P}_{2}^{(l)} \\
\mathbf{P}_{2}^{(u)}
\end{array}\right).
\end{eqnarray}
In addition, when $|x_1^+|<O(|t|^{1/2})$, in the first $k_{even}$ rows of the upper row block of the above right matrix, the $\mathbf{E}^{-1} \widehat{\mathbf{G}}\mathbf{E}_2\mathbf{P}_{2}^{(l)}$ term is subdominant to $\mathbf{P}_1$ and can be asymptotically neglected. Then, when calculating the determinant of the $\Phi_{1\le i, j\le N}$ matrix, the lower row block $\mathbf{P}_{2}^{(u)}$ with $k_{even}$ rows and those first $k_{even}$ rows in the upper row block $\mathbf{P}_1 + \mathbf{E}^{-1} \widehat{\mathbf{G}}\mathbf{E}_2\mathbf{P}_{2}^{(l)}$ all cancel out, and we get
\begin{eqnarray} \label{detPhi15}
\det_{1 \leq i, j\leq N} \Phi_{i, j} \sim \det{\cal F} \det {\cal R}_1 \det
\left( \mathbf{Y}_1 + \mathbf{E}_2^{-1} \widehat{\mathbf{G}}_2\mathbf{E}_2\mathbf{Y}_2 \right),
\end{eqnarray}
where
\[ \label{defYk}
\mathbf{Y}_k = \mbox{Mat}_{1\leq i\leq d,\ 1\leq j\leq d} \left( 2^{-(j-1)} S_{2i-j+1-k}(\textbf{v})\right),
\]
and
\[
\widehat{\mathbf{G}}_2=\mbox{Mat}_{k_{even}+1\le i \le k_{odd}, 1\le j \le d} \widehat{\mathbf{G}}
\]
is the lower $d\times d$ submatrix of $\widehat{\mathbf{G}}$. Notice that this $\mathbf{Y}_1 + \mathbf{E}_2^{-1} \widehat{\mathbf{G}}_2\mathbf{E}_2\mathbf{Y}_2$ matrix is of the same form as Eq.~(\ref{PhiAsymF}) of Sec.~\ref{secProof}. Then, following the same calculations as in the proof of Theorem~\ref{Theorem1} after Eq.~(\ref{PhiAsymF}), we can prove Theorem~\ref{Theorem2}, where $\hat{\beta}_{r,r}$ in that theorem comes from the factorization (\ref{FacAB12}) for the lower left corner submatrix of $\widehat{\mathbf{G}}$ (which is the same factorization for the lower left corner submatrix of $\widehat{\mathbf{G}}_2$ above).

\subsection{Proof for the case of $k_{even}-k_{odd}>2$}
In this case, there are more even indices than odd ones in $\Lambda$. Our starting point is still the determinant in Eq.~(\ref{3Nby3Ndet2b}). As we did in Sec.~\ref{seccase2}, we now group the even indices together in the order $n_1<n_2<\cdots<n_{k_{even}}$, followed by odd indices in the order $\hat{n}_{1}<\hat{n}_{2}< \cdots< \hat{n}_{k_{odd}}$. That is, $\Lambda=(n_1, \dots, n_{k_{even}}, \hat{n}_{1}, \dots, \hat{n}_{k_{odd}})$, with $k_{even}+k_{odd}=N$. As a consequence, the first $k_{even}$ rows of $\Phi$ above correspond to even indices, and the last $k_{odd}$ rows of $\Phi$ correspond to odd indices.

Now we redefine $T_2$-dependent matrices
\begin{eqnarray}
&& {\cal F}=\left(
             \begin{array}{cccc}
               \frac{T_2^{[n_1/2]}}{[n_1/2]!} & \frac{T_2^{[n_1/2]-1}}{([n_1/2]-1)!} & \frac{T_2^{[n_1/2]-2}}{([n_1/2]-2)!}  & \cdots \\
                \vdots &  \vdots & \vdots & \vdots \\
                \frac{T_2^{[n_{k_{even}}/2]}}{[n_{k_{even}}/2]!} & \frac{T_2^{[n_{k_{even}}/2]-1}}{([n_{k_{even}}/2]-1)!} & \frac{T_2^{[n_{k_{even}}/2]-2}}{([n_{k_{even}}/2]-2)!}& \cdots
             \end{array}
           \right)_{k_{even}\times k_{even}}, \label{defcalF5} \\
&& {\cal R}=\left(
             \begin{array}{cccc}
               \frac{T_2^{[\hat{n}_1/2]}}{[\hat{n}_1/2]!} &  \frac{T_2^{[\hat{n}_1/2]-1}}{([\hat{n}_1/2]-1)!} &  \frac{T_2^{[\hat{n}_1/2]-2}}{([\hat{n}_1/2]-2)!} & \cdots \\
                \vdots &  \vdots & \vdots & \vdots \\
               \frac{T_2^{[\hat{n}_{k_{odd}}/2]}}{([\hat{n}_{k_{odd}}/2])!} &  \frac{T_2^{[\hat{n}_{k_{odd}}/2]-1}}{([\hat{n}_{k_{odd}}/2]-1)!} & \frac{T_2^{[\hat{n}_{k_{odd}}/2]-2}}{([\hat{n}_{k_{odd}}/2]-2)!} & \cdots
             \end{array}
 \right)_{{k_{odd}}\times k_{even}},  \\
 && \mathbf{E}=\mbox{diag}\left(1, T_2^{-1}, T_2^{-2}, \cdots T_2^{-({k_{even}}-1)} \right).
\end{eqnarray}
We also split
\[
{\cal R}=\left[ {\cal R}_1, {\cal R}_2, {\cal R}_3\right], \quad
\mathbf{E}=\mbox{diag}(\mathbf{E}_0,  \mathbf{E}_1, \mathbf{E}_2),
\]
where ${\cal R}_1$ has $k_{odd}$ columns, ${\cal R}_2$ has $d$ columns, ${\cal R}_3$ has 1 column, and
\[
\mathbf{E}_0=1, \quad \mathbf{E}_1=\mbox{diag}\left(T_2^{-1}, T_2^{-2}, \cdots T_2^{-k_{odd}} \right), \quad \mathbf{E}_2=\mbox{diag}\left(T_2^{-(k_{odd}+1)}, T_2^{-(k_{odd}+2)}, \cdots T_2^{-(k_{even}-1)} \right)_{d\times d}.
\]
Matrices $({{\cal F}, \cal R}_1, {\cal R}_2, {\cal R}_3)$ are related to $(\mathbf{F}, \mathbf{R}_1, \mathbf{R}_2, \mathbf{R}_3)$ of Eqs.~(\ref{defF3}) and (\ref{defFRsplitb}) as
\[ \label{calFR2FRb}
{\cal F}=\mathbf{E}_f \mathbf{F} \mathbf{E}, \quad {\cal R}_1=T_2\mathbf{E}_r \mathbf{R}_1 \mathbf{E}_1,\ \ \ {\cal R}_{2}=T_2\mathbf{E}_r \mathbf{R}_2 \mathbf{E}_2, \quad {\cal R}_3=T_2^{-(k_{even}-1)}\mathbf{E}_r \mathbf{R}_3,
\]
where
\[
\mathbf{E}_f=\mbox{diag}\left(T_2^{[n_1/2]}, T_2^{[n_2/2]},  \cdots T_2^{[n_{k_{even}}/2]} \right), \quad
\mathbf{E}_r=\mbox{diag}\left(T_2^{[\hat{n}_1/2]}, T_2^{[\hat{n}_2/2]},  \cdots T_2^{[\hat{n}_{k_{odd}}/2]} \right).
\]

When $1\ll |x_1^+|\ll O(|t|^{1/2})$, we use relations (\ref{polyrelation3})-(\ref{polyrelation2}) and (\ref{factS1}) to write the dominant terms of $\Phi$ in Eq.~(\ref{3Nby3Ndet2b}) as
\[ \label{PhiAasym33}
\Phi_{\small{\mbox{first $k_{even}$ rows}}}  \sim {\cal F} \mathbf{P}_2+\mathbf{D}_1{\cal F} \mathbf{P}_3,
\]
\[ \label{PhiBasym33}
\Phi_{\small{\mbox{last $k_{odd}$ rows}}}  \sim {\cal R} \mathbf{P}_1+\mathbf{D}_2{\cal R} \mathbf{P}_2,
\]
where $\mathbf{P}_k$ is as defined in Eq.~(\ref{defPk123}), and
\[
\mathbf{D}_2= \mbox{diag}\left(a_{k_{even}+1,1}, a_{k_{even}+2,1}, \cdots, a_{N,1} \right).
\]
The diagonal of this $\mathbf{D}_2$ is the vector of $a_{i,1}$ parameters for the odd indices of $\Lambda$. We split matrices $\mathbf{P}_1$ and $\mathbf{P}_3$ as
\[
\mathbf{P}_1=\left(\begin{array}{c} \mathbf{P}_{1}^{(t)} \\ \mathbf{P}_{1}^{(m)} \\ \mathbf{P}_{1}^{(b)}\end{array}\right), \quad
\mathbf{P}_3=\left(\begin{array}{c} \mathbf{0} \\ \mathbf{P}_{1}^{(t)} \\ \mathbf{P}_{1}^{(m)}\end{array}\right),
\]
where $\mathbf{P}_{1}^{(t)}$ is the first $k_{odd}$ rows of $\mathbf{P}_1$, $\mathbf{P}_{1}^{(m)}$ the next $d$ rows of $\mathbf{P}_1$, $\mathbf{P}_{1}^{(b)}$ the single last row of $\mathbf{P}_1$, and
$\mathbf{0}$ a single row of zeros. Then, employing Eqs.~(\ref{defG3}), (\ref{defFRsplitb}) and (\ref{calFR2FRb}), we can rewrite the asymptotics (\ref{PhiAasym33}) and (\ref{PhiBasym33}) of $\Phi$ as
\[ \label{PhiAasym25}
\Phi_{\small{\mbox{first $k_{even}$ rows}}}  \sim {\cal F} \left(\mathbf{P}_2+\mathbf{E}^{-1}\mathbf{G} \mathbf{E}\mathbf{P}_3\right)={\cal F} \left(\mathbf{P}_2 + \mathbf{E}^{-1} \mathbf{G}_1\mathbf{E}_1\mathbf{P}_{1}^{(t)}+ \mathbf{E}^{-1} \mathbf{G}_2\mathbf{E}_2\mathbf{P}_{1}^{(m)}\right),
\]
and
\[ \label{PhiBasym25}
\Phi_{\small{\mbox{last $k_{odd}$ rows}}} \sim
{\cal R}_1\left\{
\mathbf{P}_{1}^{(t)}+ \mathbf{E}_1^{-1}\mathbf{R}_1^{-1} \mathbf{R}_2\mathbf{E}_2\mathbf{P}_{1}^{(m)}+ T_2^{-k_{even}} \mathbf{E}_1^{-1}\mathbf{R}_1^{-1}\mathbf{R}_3\mathbf{P}_{1}^{(b)}+T_2^{-1}\mathbf{E}_1^{-1}\mathbf{R}_1^{-1}\mathbf{D}_2\mathbf{R}\mathbf{E} \mathbf{P}_2\right\}.
\]
We see from Eq.~(\ref{factS1}) that the $T_2^{-k_{even}} \mathbf{E}_1^{-1}\mathbf{R}_1^{-1}\mathbf{R}_3\mathbf{P}_{1}^{(b)}$ term in Eq.~(\ref{PhiBasym25}) is subdominant compared to $\mathbf{P}_{1}^{(t)}$. Thus, we can neglect that term and get
\begin{eqnarray}
\Phi\sim \left( \begin{array}{cc}
                  {\cal F} & \textbf{O} \\
                  \textbf{O} & {\cal R}_1
                \end{array}\right)
\left(\begin{array}{l}
 \mathbf{P}_2 + \mathbf{E}^{-1} \mathbf{G}_1\mathbf{E}_1\mathbf{P}_{1}^{(t)}+ \mathbf{E}^{-1} \mathbf{G}_2\mathbf{E}_2\mathbf{P}_{1}^{(m)} \\
 \mathbf{P}_{1}^{(t)}+ \mathbf{E}_1^{-1}\mathbf{R}_1^{-1} \mathbf{R}_2\mathbf{E}_2\mathbf{P}_{1}^{(m)}+ T_2^{-1} \mathbf{E}_1^{-1}\mathbf{R}_1^{-1}\mathbf{D}_2\mathbf{R}\mathbf{E} \mathbf{P}_2
\end{array}\right).
\end{eqnarray}

Now, we use the lower $\mathbf{P}_{1}^{(t)}$ block of the above right matrix to eliminate the upper $\mathbf{E}^{-1} \mathbf{G}_1\mathbf{E}_1\mathbf{P}_{1}^{(t)}$ term of that matrix through row operations, which does not affect the $\sigma$ determinant in Eq.~(\ref{3Nby3Ndet2b}). From this step we get
\[
\Phi\sim \left( \begin{array}{cc}
                  {\cal F} & \textbf{O} \\
                  \textbf{O} & {\cal R}_1
                \end{array}\right) \left(\begin{array}{l}
 \mathbf{P}_2 + \mathbf{E}^{-1} \widehat{\mathbf{G}}\mathbf{E}_2\mathbf{P}_{1}^{(m)}-T_2^{-1}
 \mathbf{E}^{-1}\mathbf{G}_1\mathbf{R}_1^{-1}\mathbf{D}_2\mathbf{R}\mathbf{E}\mathbf{P}_2 \\
\mathbf{P}_{1}^{(t)}+ \mathbf{E}_1^{-1}\mathbf{R}_1^{-1} \mathbf{R}_2\mathbf{E}_2\mathbf{P}_{1}^{(m)}+ T_2^{-1} \mathbf{E}_1^{-1}\mathbf{R}_1^{-1}\mathbf{D}_2\mathbf{R}\mathbf{E} \mathbf{P}_2
\end{array}\right),
\]
where $\widehat{\mathbf{G}}$ is as defined in Eq.~(\ref{defGhat2}).

Following similar steps as before, the above asymptotics can be reduced to
\begin{eqnarray}
\Phi\sim \left( \begin{array}{cc}
                  {\cal F} & \textbf{O} \\
                  \textbf{O} & {\cal R}_1
                \end{array}\right) \left(\begin{array}{l}
 \mathbf{P}_2 + \mathbf{E}^{-1} \widehat{\mathbf{G}}\mathbf{E}_2\mathbf{P}_{1}^{(m)} \\
\mathbf{P}_{1}^{(t)}
\end{array}\right).
\end{eqnarray}
In addition, when $|x_1^+|<O(|t|^{1/2})$, in the first $k_{odd}+1$ rows of the upper row block of the above right matrix, the $\mathbf{E}^{-1} \widehat{\mathbf{G}}\mathbf{E}_2\mathbf{P}_{1}^{(m)}$ term is subdominant to $\mathbf{P}_2$ and can be asymptotically neglected. Then, when calculating the determinant of the $\Phi_{1\le i, j\le N}$ matrix, the lower row block $\mathbf{P}_{1}^{(t)}$ of $k_{odd}$ rows and those first $k_{odd}+1$ rows in the upper row block of the above right matrix all cancel out, and we get
\begin{eqnarray}
\det_{1 \leq i, j\leq N} \Phi_{i, j} \sim \det{\cal F} \det {\cal R}_1 \det
\left( \mathbf{Y}_1 + \mathbf{E}_2^{-1} \widehat{\mathbf{G}}_2\mathbf{E}_2\mathbf{Y}_2 \right),
\end{eqnarray}
where $\mathbf{Y}_k$ is as defined in Eq.~(\ref{defYk}), and
\[
\widehat{\mathbf{G}}_2=\mbox{Mat}_{k_{odd}+2\le i \le k_{even}, 1\le j \le d} \widehat{\mathbf{G}}
\]
is the lower $d\times d$ submatrix of $\widehat{\mathbf{G}}$. The above $\mathbf{Y}_1 + \mathbf{E}_2^{-1} \widehat{\mathbf{G}}_2\mathbf{E}_2\mathbf{Y}_2$ matrix is of the same form as Eq.~(\ref{PhiAsymF}) of Sec.~\ref{secProof}. Then, following the same calculations as in the proof of Theorem~\ref{Theorem1} after Eq.~(\ref{PhiAsymF}), we can prove Theorem~\ref{Theorem2}, where $\hat{\beta}_{r,r}$ in that theorem comes from the factorization (\ref{FacAB12}) for a submatrix of $\widehat{\mathbf{G}}$.

\section{Summary and Discussion} \label{sec7}

In this article, we have analytically studied large-time patterns of general higher-order lump solutions in the KP-I equation. We have shown that when the index vector of the general lump solution is a sequence of consecutive odd integers starting from one, the large-time pattern generically would comprise fundamental lumps uniformly distributed on concentric rings. In addition, the fundamental lumps on these rings separate from each other in proportion to $|t|^{m/(2m+1)}$, where $m$ is a positive integer that takes on different values on different rings. For other index vectors, we have shown that the large-time pattern of a general higher-order lump would comprise fundamental lumps in the outer region as described analytically by the nonzero-root structure of the associated Wronskian-Hermit polynomial, together with possible fundamental lumps in the inner region that are uniformly distributed on concentric rings generically. Leading-order predictions of fundamental lumps in these solution patterns have also been derived. Our predicted patterns at large times have been compared to true solutions, and good agreement has been obtained.

Earlier in \cite{YangYangKPI}, large-time patterns of special higher-order lump solutions in the KP-I equation were determined. Those lump solutions were special because their internal parameters $a_{i, j}$ were required to be independent of the $i$ index. By comparing the results of this paper with those in \cite{YangYangKPI}, the biggest difference is that the triangular patterns as reported in \cite{YangYangKPI} for those special higher-order lumps are now replaced by concentric-ring patterns for generic general higher-order lumps. This big difference in wave patterns for special and generic general cases is a surprise.

A mathematical difference between the special case of \cite{YangYangKPI} and the current generic general case is that, the triangular pattern in \cite{YangYangKPI} was described analytically by the root structure of a certain Yablonskii-Vorob'ev polynomial, but the concentric-ring pattern here is described analytically by the root structure of a simpler two-term polynomial in Eq.~(\ref{Qnz}). Roots of the Yablonskii-Vorob'ev polynomial do not have explicit formulae, but roots of that two-term polynomial in (\ref{Qnz}) do. Thus, the polynomials used for the prediction of lump patterns are simpler in the generic general case than in the special case. This mathematical difference is also a surprise.

We emphasize that these concentric-ring lump patterns would appear for generic internal parameters of higher-order lumps which meet Assumption 1 or 2 of this paper. If internal parameters do not meet such generic assumptions, large-time solution patterns would be different. One known example is the special parameters in \cite{YangYangKPI} which do not meet such assumptions, and the corresponding lump patterns are triangular instead of concentric rings. As another example, suppose the index vector is $\Lambda=(1, 3, 5, 7, 9)$. Assumption 1 for this case requires $M_1\ne 0$ and $M_2\ne 0$. If our internal parameters only meet the former condition of $M_1\ne 0$ but not the latter condition of $M_{2}\ne 0$, i.e., if $M_1\ne 0$ but $M_2=0$ now, then the large-time pattern of the solution would not be concentric rings as shown in Fig.~\ref{f:fig2}, but a single ring of nine fundamental lumps plus a triangle of six fundamental lumps inside the ring. For these nongeneric internal parameters, their large-time solution patterns can also be asymptotically predicted using techniques of this paper, but this will not be pursued in this article.

\section*{Acknowledgment}
The work of B.Y. was supported in part by the National Natural Science Foundation of China (Grant No. 12201326), and the work of J.Y. was supported in part by the National Science Foundation (U.S.) under award number DMS-1910282.

\end{document}